\title{Reynolds number limits for jet propulsion: A numerical study of simplified jellyfish}
\author{Gregory Herschlag and Laura A. Miller}
\date{\today}

\documentclass[12pt]{article}

\usepackage{amsmath,amssymb,latexsym,amsthm}
\usepackage[top=1in, bottom=1in, left=1in, right=1in]{geometry}
\usepackage[pdftex]{graphicx}

\begin{document}
\pagestyle{empty}
\begin{flushleft}
\begin{tabular}{ll}
\hline
Title \hspace{1.4in} & Reynolds number limits for jet propulsion: A numerical study \\
&  of simplified jellyfish\\
 \hline
\end{tabular}
\end{flushleft}
\begin{tabular*}{6.5in}{lll}
 Corresponding Author \hspace{.13in} & Family Name \hspace{.7in} & Miller \hspace{1in}\\
& Given Name & Laura\\
& Division & Department of Mathematics\\
& Organization & University of North Carolina\\
& Address & 27599-3250, Chapel Hill, NC, USA\\
& Email & lam9@email.unc.edu\\
\hline
Author \hspace{.13in} & Family Name \hspace{.7in} & Herschlag      \\
& Given Name & Gregory\\
& Division & Department of Mathematics\\
& Organization & University of North Carolina\\
& Address & 27599-3250, Chapel Hill, NC, USA\\
& Email & gregoryh@email.unc.edu\\
\hline
 \end{tabular*}
  \\
  \newline
   \underline{Abstract}
  \\
The Scallop Theorem states that reciprocal methods of locomotion, such as jet propulsion or paddling, will not work in Stokes flow (Reynolds number = 0). In nature the effective limit of jet propulsion is still in the range where inertial forces are significant. It appears that almost all animals that use jet propulsion swim at Reynolds numbers (Re) of about 5 or more. Juvenile squid and octopods hatch from the egg already swimming in this inertial regime. Juvenile jellyfish, or ephyrae, break off from polyps swimming at $Re$ greater than 5.  Many other organisms, such as scallops, rarely swim at $Re$ less than 100. The limitations of jet propulsion at intermediate $Re$ is explored here using the immersed boundary method to solve the two-dimensional Navier Stokes equations coupled to the motion of a simplified jellyfish. The contraction and expansion kinematics are prescribed, but the forward and backward swimming motions of the idealized jellyfish are emergent properties determined by the resulting fluid dynamics. Simulations are performed for both an oblate bell shape using a paddling mode of swimming and a prolate bell shape using jet propulsion. Average forward velocities and work put into the system are calculated for $Re$ between 1 and 320. The results show that forward velocities rapidly decay with decreasing $Re$ for all bell shapes when $Re<10$. Similarly, the work required to generate the pulsing motion increases significantly for $Re<10$. When compared actual organisms, the swimming velocities and vortex separation patterns for the model prolate agree with those observed in \textit{Nemopsis bachei}. The forward swimming velocities of the model oblate jellyfish after two pulse cycles are comparable to those reported for \textit{Aurelia aurita}, but discrepancies are observed in the vortex dynamics between when the 2D model oblate jellyfish and the organism.  This discrepancy is likely due to a combination of the differences between the 3D reality of the jellyfish verses the 2D simplification, as well as  the rigidity of the time varying geometry imposed by the idealized model.
  \\
  \newline
   \newline
   \underline{Notes}
  \\
  All figures that are submitted in color should be presented in color where possible

\section{Introduction}
Methods of effective locomotion must utilize and contend with both viscous and inertial forces throughout their execution.  The ratio between these two forces (inertial forces divided by viscous forces) is famously known as the Reynolds number (\textit{Re}). \textit{Re} is often used when discussing scaling effects in fluid dynamics, as systems of the same \textit{Re} are dynamically similar. Low \textit{Re} flows are reversible, and a consequence of reversibility is that net fluid transport and locomotion do not occur by reciprocal motions. This result is known as the Scallop Theorem which was first introduced by~\cite{Purc:77}. One of the implications of this theorem is that scallop jet propulsion is not possible at $Re=0$. The theorem gets its name from an idealized scallop which is shown to move forward upon contraction of its shell but then slides back to its original position upon opening its shell. Other mechanisms of reciprocal locomotion such as pectoral fin swimming in fish and flapping in stingrays are also not possible at very low \textit{Re}. High \textit{Re} flows, on the other hand, are dominated by pressure and inertial forces, and locomotion is possible using reciprocal motions such as flapping and undulating.

In the natural world, it appears that organisms do not use reciprocal methods of locomotion below $Re$ of $\mathcal{O}({1})$.  There are a number of ways to calculate $Re$, but for the purpose of the following discussion we will use the following definitions:
\begin{equation}
Re_m = \frac{\rho U_{ave} D}{\mu}
\end{equation}
and
\begin{equation}
Re_k = \frac{\rho U_{body} D}{\mu}
\end{equation}
where $Re_m$ is the movement based Reynolds number,  $Re_k$ is the kinematic based Reynolds number, $\rho$ is the density of the fluid, $U_{ave}$ is the average forward swimming velocity, $U_{body}$ is a characteristic speed of the swimmer with respect to itself, $D$ is some characteristic length of the organism, and $\mu$ is the dynamic viscosity of the fluid. Flapping fins and undulatory swimming do not appear for $Re_m<10$~\cite{Alben:05,Childress:04, Vanden:04}. Similar physical limits also appear to exist for jet propulsion. Both squid and octopus species hatch from the eggs already swimming at $Re_m > 10$ and grow to higher regimes (calculated from~\cite{Thom:01, Bole:01}). Juvenile scallops use jet propulsion at $Re_m > 200$ with peak performance at $Re_m > 3000$~\cite{Manuel:93} 
In jellyfish, ephyrae  break off from polyps and swim at $Re_m > 1$ ~\cite{Feitl:09}. Ephyrae continue to grow into the mature medusae that swim at $Re_m>100$~\cite{Higg:08, Feitl:09}. 
While many studies have considered jellyfish swimming at $Re > 100$, it has yet to be studied how methods of reciprocal swimming break down as viscous forces steadily increase.

In this work, an idealized model of jellyfish swimming is used to explore the effects of $Re$ on jet propulsion. The relatively simple design of the jellyfish bell makes it well suited for fluid-structure interaction (FSI) studies. Several research groups have previously used FSI finite difference and finite element methods to study the flows generated by hemielliptical jellyfish bells~\cite{Cure:08, Huan:09, Zhao:08}. In these these cases, the bells move with a prescribed motion so that forward velocity is an input into the system. Moshseni and Sahin ~\cite{Mohs:09} used a Lagrangian-Eulerian formulation to numerically solve for emergent forward motion in the jellyfish. Actual bell profiles were used as inputs into the simulations. In this paper, the immersed boundary method is used to solve the FSI problem for a two-dimensional jellyfish. Preferred contraction and expansion kinematics are used as inputs for the simulations, but the forward motion of the jellyfish is due to the resulting fluid motion. Given the computational demands of solving the FSI problem for a wide parameter space, simplified bell shapes and kinematics in 2D flows are used so that the study is tractable. With this in mind, the purpose of this study is to explore how scale and morphology affect forward locomotion velocities in jet propulsion rather than to accurately simulate the flows generated by specific medusae.

To consider more than one form of jet propulsion, immersed boundary simulations are performed for oblate and prolate medusae and hyperexpanding ephyrae. The resulting flow structures produced by the computational medusae are then compared to those measured in actual organisms. In all cases, rhythmic contractions drive the fluid out of the bell and simultaneously move the animal forward. Depending upon the shape of the bell, the swimming mechanism can be classified as either rowing or true jet propulsion. The rowing or paddling mode is found in oblate medusae such as \textit{Aurelia aurita}~\cite{Dabi:05}. A paddling-type mechanism of swimming is also used by the ephyrae.  Although the ephyral form is a discontinuous surface with deep clefts, viscous effects  prevent flow between the lappets and create a hydrodynamically continuous surface~\cite{Nawroth:10,Feitl:09}. As a result, the ephyrae swim with a paddling motion reminiscent of oblate medusae. Prolate hydromedusae, such as \textit{Nemopsis bachei}~\cite{Dab:06}, are described as using true jet propulsion. For each body form, work put into the system and average swimming velocities are calculated for a collection of $Re_k$ ranging from 1 to 320.  For the remainder of the paper, $Re$ will refer to $Re_k$.

\section{Methods}\label{methods}
\subsection{Simplified Jellyfish Model}
The numerical simulations are constructed such that a 2D plane cuts through the axis of symmetry of the jellyfish, generating a cross section of the bell with maximum diameter. This cross section is then modeled as an ellipse which is erased below some lower bound.  The idea of approximating a jellyfish as a hemiellipsoid, while an idealization, has been used by Colin and Costello~\cite{Coli:02}, Daniel~\cite{Dani:85}, and McHenry and Jed~\cite{McHenry:03}.  The additional degree of freedom of cutting the ellipsoid anywhere along the horizontal axis is used to better capture the geometry of the jellyfish before and after contraction.

\begin{figure}[htp]
\centering
\scalebox{.5}{\includegraphics{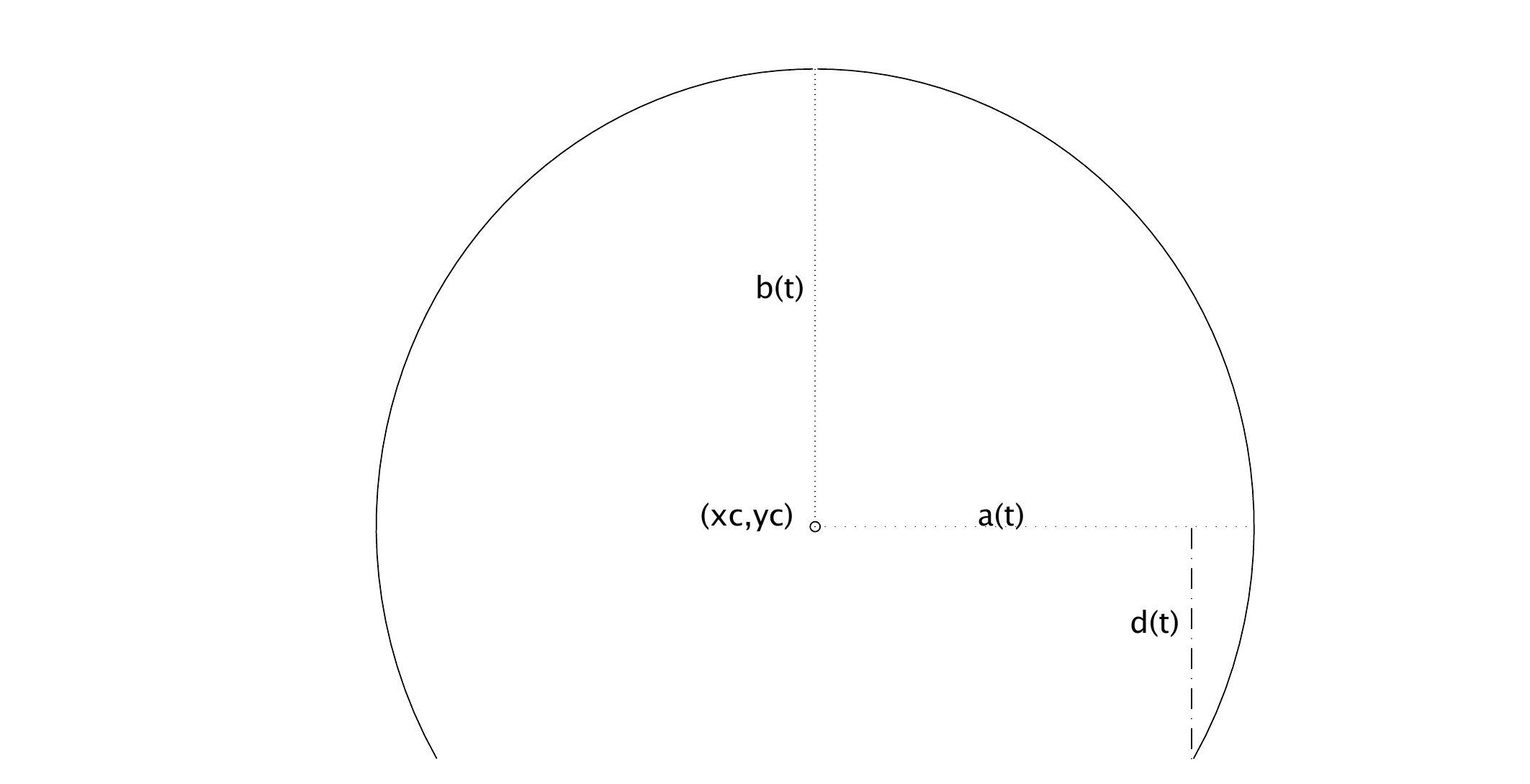}}
\caption{Diagram of the idealized jellyfish cross section. $a(t)$ gives the half width of the bell as a function of time, $b(t)+d(t)$ gives the length of the bell as a function of time, and $(x_c,y_c)$ gives the center of the ellipse that is used to make the bell.}
\label{fig:belldiag}
\end{figure}

At a given instance in time, the geometry of the bell is described using the terms $a(t)$, $b(t)$, $d(t)$, $(x_c,y_c)$ (Figure~\ref{fig:belldiag}). The shape is then given by
\begin{equation}\label{shapeeqn}
1 = \frac{(x-x_c)^2}{a(t)^2} + \frac{(y-y_c)^2}{b(t)^2} \text{  for  } y\geq y_c-d(t)
\end{equation}
where $a(t)$ is the half-width of the bell as a function of time, $b(t)+d(t)$ is the height of the bell as a function of time, $x_c$ is the centerline of motion which remains fixed, and $y_c$ is determined by the motion of the fluid.  In order to obtain geometries for both the completely relaxed and contracted oblate medusae, the above parameters were estimated from Figure 5 of Dabiri \textit{et al.}~\cite{Dabi:05} using least squares.  The values obtained from this analysis are listed in Table \ref{param}.  The parameter values for ephyra and prolate jellyfish are found by adjusting the values from the oblate data to generate reasonable shapes.

\begin{table}
\begin{tabular}{l*{7}{c}r}
Parameter             & $a_i$ & $b_i$ & $d_i$ & $d_f$ &$p_a$ & $p_b$ & $p_d$ & $t_c$ \\
\hline
Real Oblate & .051 & 0.0211 & 0.001759 & - & 0.3103 & 0.25 & -3.0 & .63  \\
Ephyra    & .031 & -.01 & 0.001759 & 0.001758 & .51 & 2.5 & - & .63 \\
Prolate     & .05 & 0.075 & 0.025 & - & .5 &  .2 & 0 &  2.43  \\
\end{tabular}
\caption{In all runs the time to refill ($t_r$) is three times $t_c$.  Each run was swept over $Re = 2^n$ with $n = (0,1,2,3,4,5,6,7)$ (the prolate was also run at 160 and the oblate was also run at 96, 160, 320) where the kinematic viscosity was calculated from Equations \ref{re}, \ref{char-l} and \ref{char-U}.}
\label{param}
\end{table}


To move the jellyfish bell from contracted to expanded states, time is parameterized so that $s = 0$ corresponds to a completely relaxed state, and $s = 1$ corresponds to a completely contracted state.  The equations giving the values of $a$ and $b$ as functions of time during the contraction are defined as

\begin{equation}\label{kineticsa}
a(s) = a_i (1-s(p_a))
\end{equation}
\begin{equation}\label{kineticsb}
b(s) = b_i (1-s(p_b))
\end{equation}
 where $a_i$ is the initial half width of the bell and $b_i$ is the initial height of the top part of the bell. For the cases of the oblate and prolate jellyfish, the equation for $d$ is defined as
 \begin{equation}\label{kineticsd}
d(s) = d_i (1-s(p_d)).
\end{equation}

 In the above equations, $b_i+d_i$ is the initial height of the bell, $p_a$ is the percentage of contraction of the half width of the bell, $p_b$ and $p_d$ are the percentages of change in $b(s)$ and $d(s)$ respectively. The equations for the expansion kinematics are constructed similarly. $s$ is scaled to account for the difference in contraction and expansion times by the equation
 \begin{equation}\label{kineticssm}
    s = \frac{1}{2}(1+\sin(\frac{\pi (2\tau-1)}{2}))
\end{equation}
 where $\tau$ is the time shifted and scaled to run linearly with respect to the real time from $0$ to $1$. $s$ was chosen to vary smoothly in time like a $\sin$ function, roughly approximating the velar diameters measured by Dabiri \textit{et al.}~\cite{Dab:06}. $\tau$ is defined as
\begin{equation}\label{kinetic_cont}
 \tau = \frac{t-t_0}{t_c} \text{  (for contraction)}
\end{equation}
\begin{equation}\label{kinetic_exp}
 \tau = 1-\frac{t-t_0}{t_r} \text{  (for expansion)}
\end{equation}
where $t_0$ is the start time of either a contraction or expansion, $t_c$ is the time it takes to contract, and $t_r$ is the time it takes to refill.

The time of contraction for the oblate jellyfish is set to $t_c=0.63$ seconds, and the time of refilling was set to $t_r=3 t_c$ seconds.  These numbers reflect the real time of contraction of the oblate jellyfish, \emph{Aurelia aurita}~\cite{Dabi:05}. The contraction time of the prolate jellyfish was set to $t_{c}=2.43$ seconds, and time of refilling was set to $t_r = 3 t_{c}$. This description gives the shape and horizontal position of the jellyfish at any instant in time.

Ephyral bells are not shaped like a continuous disk but rather have deep clefts between the lappets. Flow visualization studies~\cite{Nawroth:10, Feitl:09} suggest that there is little flow through the clefts due to viscous effects at these low $Re$. As a result, the surface of the ephyral bell acts as a hydrodynamically continuous surface. The ephyra are also able to hyperextend during the expansion of the bell due to the presence of the clefts. To approximate this motion, the equation for $d$ is constructed as follows:
\begin{equation}
   d(s) = \left\{
     \begin{array}{lr}
       d_i((\frac{1}{p_b}-s)p_b)^3 &  :  s < \frac{1}{p_b}\\
       d_f(\frac{p_b}{p_b-1}(s-\frac{1}{p_b}))^3 & :  s \geq \frac{1}{p_b}
     \end{array}
   \right.   \text{\\   (for the ehpyra)}
\end{equation}
The choice of $d(s)$ was made to ensure that the body is flat when $b(s) = 0$. The times of contraction and refilling are the same as for the oblate case ($t_c=0.63$ seconds and $t_r=3 t_c$ seconds).

\subsection{Numerical Method}
To solve the fluid-structure interaction problem, a two-dimensional version of the immersed boundary method is used~\cite{Peskin02}. The immersed boundary method has been applied to a wide variety of problems in biological fluid dynamics for intermediate $Re$ including insect flight~\cite{Miller:04, Miller:05, Miller:09}, aquatic locomotion~\cite{Fauci88,Fauci93}, and ciliary driven flows~\cite{Grun:98}. These problems typically involve a flexible structure immersed in an incompressible fluid.

The equations of motion for the two-dimensional, incompressible fluid are
\begin{equation}\label{n-s}
\rho (\frac{\partial \textbf{u}(\textbf{x}, t)}{\partial{t}}+\textbf{u}(\textbf{x}, t)\cdot \nabla \textbf{u}(\textbf{x},t)) = -\nabla p(\textbf{x}, t) + \mu \triangle \textbf{u}(\textbf{x}, t) + \textbf{F}(\textbf{x},t)
\end{equation}
and
\begin{equation}\label{compress}
\nabla \cdot \textbf{u}(\textbf{x},t) = 0
\end{equation}
where $\textbf{u}(\textbf{x},t)$ is the fluid velocity, $p(\textbf{x},t)$ is the pressure, $\textbf{F}(\textbf{x},t)$ is the force per unit area applied to the fluid by the immersed body. The independent variables are the position vector $\textbf{x} = (x,y)$ and the time $t$.

The interaction between the fluid and the boundary is described by
\begin{equation}\label{forceex}
\textbf{F}(\textbf{x},t) = \int \textbf{f}(r,t) \delta
(\textbf{x}-\textbf{X}(r,t)) dr
\end{equation}
and
\begin{equation}\label{veex}
\frac{\partial\textbf{X}(r,t)}{\partial t} = \textbf{U}(\textbf{X}(r,t)) = \int \textbf{u}(\textbf{x},t) \delta(\textbf{x} - \textbf{X}(r,t)) d\textbf{x}
\end{equation}
where $\textbf{f}(r,t)$ is the force per unit length applied to the body as a function of Lagrangian position $r$ and time $t$, $\delta(\textbf{x})$ is a two-dimensional delta function, and $\textbf{X}(r,t)$ gives the Cartesian coordinates at time $t$ of the material point labeled by the Lagrangian parameter $r$.  Equation \ref{forceex} describes how the force is spread from the boundary to the fluid.  Equation \ref{veex} evaluates the local velocity of the fluid at the boundary.  In the numerical scheme the boundary is moved at the local fluid velocity at each time step which enforces the no-slip condition.  Each of these equations involves a two-dimensional delta distribution $\delta$ that acts as the kernel of an integral transformation.  These equations convert Lagrangian variables to Eulerian variables and \textit{vice versa}.

The basic idea behind the implementation of the numerical method is as follows:
\begin{enumerate}
\item At each time step, calculate the forces the boundaries impose on the fluid. These forces are determined by the elastic spring forces connecting the boundary to the target and pairs of boundary points to each other.
\item Spread the force from the Lagrangian grid describing the position of the boundaries to the Cartesian grid used to solve the Navier-Stokes equations (equation \ref{forceex}).
\item Solve the Navier-Stokes equations for one time step.
\item Use the new velocity field to update the position of the boundary. The boundary is moved at the local fluid velocity, enforcing the no-slip condition (equation \ref{veex}).
\end{enumerate}
For the details of the exact discretization of the immersed boundary method used here, please see Peskin~\cite{Peskin96}.

\subsection{Discretization and Structure of the Boundary}

The numerical jellyfish is designed to move forward or backwards freely along the y-axis with preferred contraction and expansion kinematics along the x-axis. In the immersed boundary method, the interface is not represented explicitly as a boundary but rather as a singular force acting on the fluid. To implement this method with preferred contraction and expansion kinematics, one must specify a singular force rather than the position of the boundary. The force required is generally not known a priori. One way to estimate this force is to compare the location of the boundary to its desired position at each time step. A force is then applied that is proportional to this difference. The error between the actual and desired motion of the boundary is then controlled by this constant of proportionality~\cite{Miller:04}.

The immersed boundary is discretized so that at least two node points from the Lagrangian grid lie inside a box made from the Cartesian grid.  This ensures that as the body moves through the fluid, no fluid will leak through the boundary to the other side.  To discretize this particular shape we choose the parameterization:
\begin{eqnarray}\nonumber \label{shape1}
(\pm a(t)\sqrt{1-(\frac{((b(t)+d(t)-\epsilon)\sin(\frac{\pi r_i}{2})-d(t)}{b(t)})^2}+x_c, \\
(b(t)+d(t)-\epsilon)\sin(\frac{\pi r_i}{2})-d(t)+y_c)
\end{eqnarray}
where $\epsilon$ is taken to be a small parameter that allows for an even number of node points in the Lagrangian discretization , and $r_i$ is a discretized version of $r$ from equations \ref{forceex} and \ref{veex} that varies from 0 to 1 ($r_i = \frac{i}{N}$ where $N$ is the number of node points). Notice that if $a(t) = b(t)$, one would obtain an equipartitioned circle. As long as the two values are reasonably close to one another, the discretization should be close to being equipartitioned (with the final requirement of choosing $\epsilon$ intelligently).  In the ephyral case the body flattens for a part of its motion; here the grid still meets the spacing requirements, however there are many more node points at the end of the body than the middle.


To prescribe contraction and expansion kinematics while allowing the bell to move forward and backward freely, the equations that describe the force the boundary applies to the fluid are distinct in $x$ and $y$.  For the $x$ component, a `target boundary' method is used. In the $y$ direction, deviations from the desired shape of the bell are penalized.

\begin{figure}
\centering
\scalebox{.35}{\includegraphics{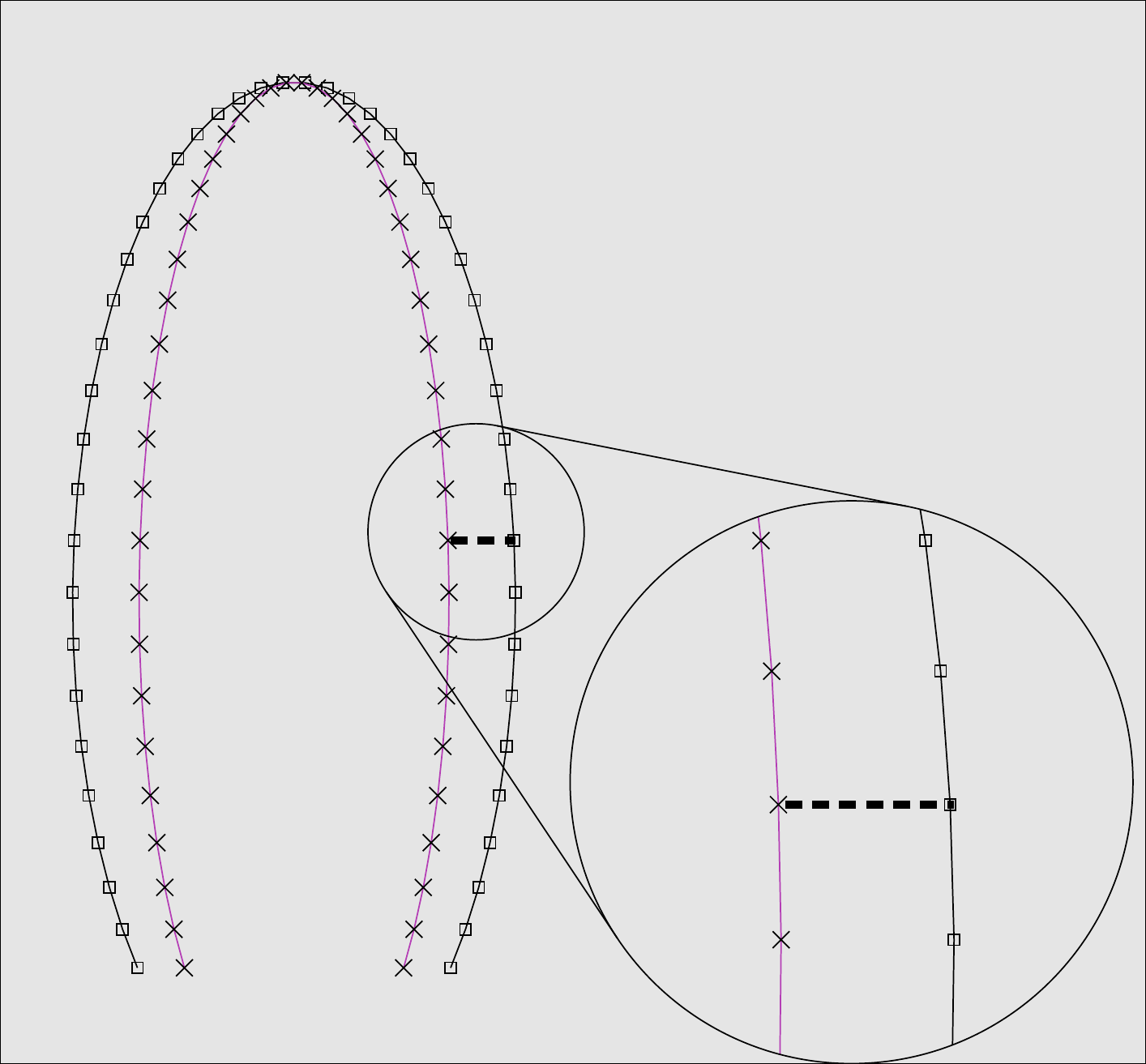}}
\caption{Diagram of the target boundary (x) and actual boundary (square). The zoomed in view shows the linear springs that connect both boundaries. In the actual simulation, springs are placed between each target and its corresponding boundary point. The distance between the two boundaries has been enlarged for clarity.}
\label{fig:force}
\end{figure}

\begin{figure}
\centering
\scalebox{.35}{\includegraphics{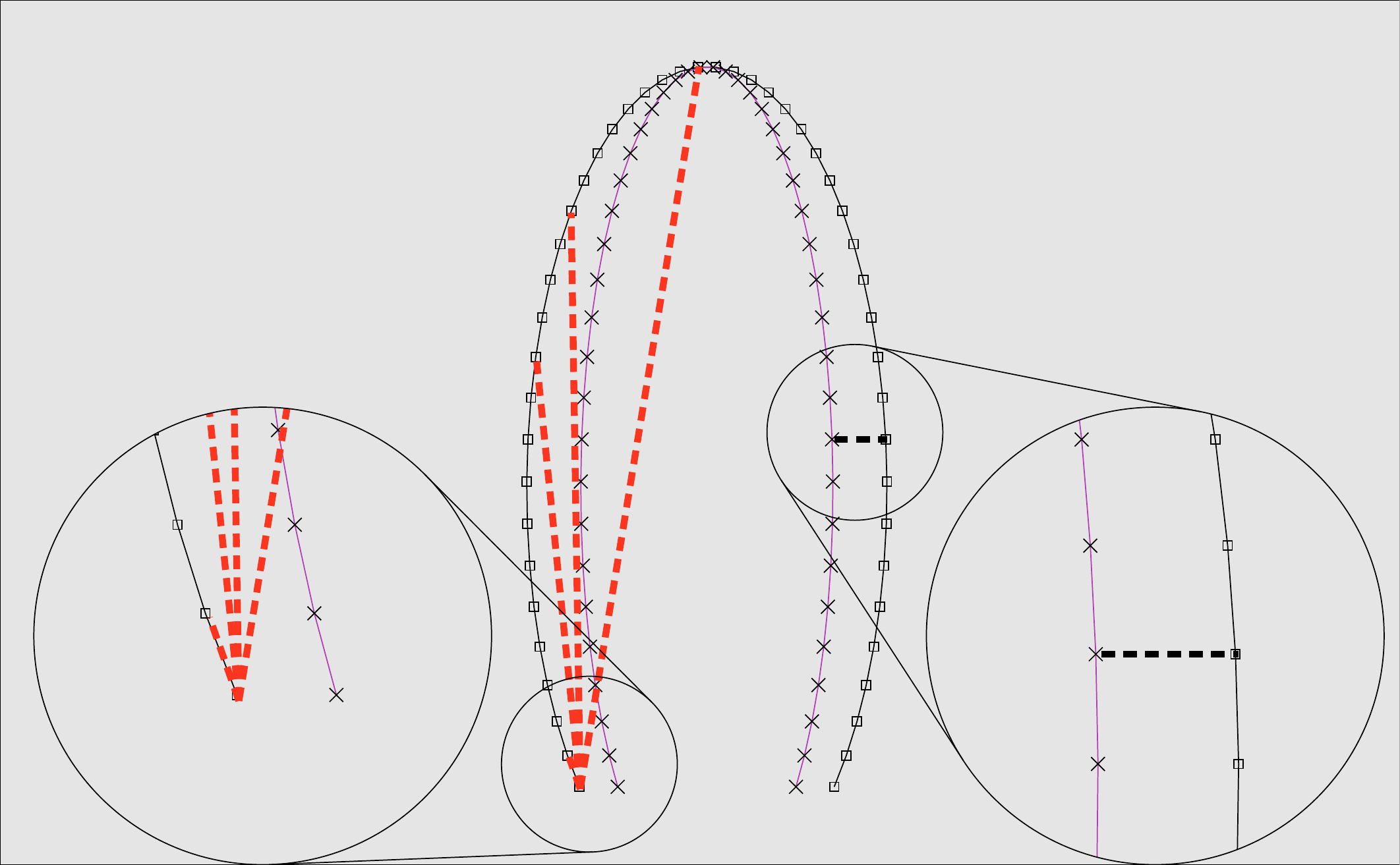}}
\caption{Diagram showing the forces used in both the x- and y-directions. The force in the y-direction is used to maintain the structure of the bell while allowing it to move freely forwards and backwards. These springs (shown by the red dashed lines) are placed between boundary points to ensure that their distances in the y-direction are preserved.}
\label{fig:f2}
\end{figure}

For the target boundary method, a boundary that does not interact with the fluid is attached with virtual springs to the actual immersed boundary (Figures~\ref{fig:force} and~\ref{fig:f2}).  The target boundary moves with the desired motion, and the springs are incorporated as a force in the x-direction that is linearly proportional to the distance between the target points and the actual points. This force is then spread to the Cartesian grid where the Navier-Stokes equations are solved.

To ensure that the bell keeps its shape during expansion and contraction, bending and stretching stiffness is enforced by connecting sets of adjacent points on the discretized Lagrangian grid with linear springs (see Figure~\ref{fig:f2}).  The number of node points, $N$, is chosen to be divisible by a set of small primes.  The particular value used for these simulations was chosen to be $2^3\times3\times5\times7$. Given any node, it is connected in the $y$ component to any adjacent node and to nodes that are $N/2$, $N/3$, $N/4$, $N/5$, $N/7$ and $N/8$ node points away on the Lagrangian grid. If two points that are connected move away from one another with a displacement in $y$ different from that prescribed by equation \ref{shape1}, a corresponding force is applied to return the distance to equilibrium.  Linking the points in this way ensures that the shape of the bell is maintained while allowing the body to slide in the $y$ direction.

Summarizing the above description leaves the following formula for $\textbf{f}(r,t)$:
\begin{eqnarray}\nonumber
\textbf{f}({r_i},t) &=& k_{s_1}[\textbf{T}({r_i},t) - \textbf{X}({r_i},t)]\cdot e_1  \\
\nonumber
&& + \sum_{{r_j}\diamondsuit {r_i}} k_{s_{({r_j},{r_i})}}\text{sgn} +((\textbf{X}({r_j},t)-\textbf{X}({r_i},t))\cdot e_2) ([\textbf{X}({r_i},t) - \textbf{X}({r_j},t)] \\
&& - [\textbf{T}({r_i},t) - \textbf{T}({r_j},t)])\cdot e_2
\end{eqnarray}
where $k_{s_1}$ is the spring coefficient between the target points and the immersed boundary, ${r_j}\diamondsuit {r_i}$ means $r_j$ is connected to $r_i$, $\textbf{T}(r,t)$ describes the location of the target points, $e_1 = (1,0)$ and $e_2=(0,1)$ are the standard Cartesian basis vectors, sgn returns the sign of its argument, and $k_{s_{({r_i,r_j})}} = k_{s_{({r_j,r_i})}}$ is the spring coefficient connecting the $y$ components of two nodes.  

Once the force on each of the Lagrangian points is determined, it is then spread to the Cartesian grid.  This is done by discretizing equation \ref{forceex} and using the approximation to the delta function described in~\cite{Minion:00}.  Discretizing equation \ref{forceex} gives

\begin{equation}
\textbf{F}_{i,j} = \sum_{k=1}^{N} \textbf{f}_k D_h(\textbf{x}_{i,j}-\textbf{X}_{r_k})\Delta l
\end{equation}

\noindent where $\textbf{F}_{i,j}$ is the force on the Cartesian grid at the node labeled $(i,j)$, $\textbf{f}_k = \textbf{f}(r_k, t)$, $\textbf{x}_{i,j}$ gives the Cartesian coordinates of the node labeled $(i,j)$, $h$ is the spatial step size of the Cartesian grid,
$\textbf{X}_{r_k}$ represents the coordinates of $r_k$ on the Cartesian grid, and $\Delta l$ is spatial step size on the Lagrangian grid approximated by $h/2$. Finally, $D_h(\textbf{x}) = d_h(x) d_h(y)$, where $d_h(x)$ is the approximation of the delta distribution. The choice of $d_h(x)$ and is chosen to be

\begin{equation}
d_h(x) = \left\{
     \begin{array}{lr}
       \frac{1}{4h}(1+\cos(\frac{\pi x}{2 h})) & , |x| \leq 2h\\
       0 & , |x| > 2h.
     \end{array}
   \right.
\end{equation}

Scaling effects are studied by varying the the kinematic viscosity of the system. For this parametric study, the $Re$ is defined (as mentioned in the introduction) so that the characteristic velocity is an input into the simulation rather than an emergent property. This $Re$ is defined as
\begin{equation}\label{re}
Re = Re_k = \frac{\rho l U_{body}}{\mu} = \frac{l U_{body}}{\nu}
\end{equation}
where $\nu$ is the kinematic viscosity, $U_{body}$ is a characteristic velocity calculated from the contraction of the bell, and $l$ is the diameter of the jellyfish.  The advantage of this formulation is that the $Re$ is uncoupled from the forward velocity, and organisms with no net forward motion are not necessarily pulsing at $Re=0$. $U_{body}$ and $l$ are given by the equations
\begin{equation}\label{char-l}
l = a_i
\end{equation}
\begin{equation}\label{char-U}
U_{body} = \frac{\sqrt{(a_ip_a)^2 + (b_ip_b)^2}}{t_c}.
\end{equation}
$U_{body}$ gives an estimate of the average tip velocity during contraction.

In the simulations that follow, \textit{Re} is varied by changing the kinematic viscosity of the fluid. We consider $Re = 2^n$ with $n=(0,1,2,3,4,5,6,7)$, with the addition of $Re = 96$, $160$ and 320 for the oblate case, and 160 for the prolate case.

The system of differential and integral equations given by the above equations was solved on a rectangular grid with periodic boundary conditions in both directions as described by Peskin and McQueen~\cite{Peskin96}.  The velocity near the outer boundary of the domain was kept near zero on the edges of the domain by inserting four walls that were 4 grid steps away from the edges of the fluid domain.  The Navier-Stokes equations were discretized on a fixed Eulerian grid.  For all oblate and prolate runs the Cartesian domain was a $512\times512$ mesh, with the length and width scales equal to $0.25\times0.25$ meters.
For the prolate jellyfish, the system was solved on a $512\times1024$ mesh with a domain width and height of $0.5\times1$ meters, with the exception of the faster cases where the mesh was changed to $512\times1536$ meters and $0.5\times1.5$ meters ($Re = 64, 128$) to accommodate for the extra distance traveled.  For the prolate case of $Re=160$ the length scales were still $0.5\times1.5$ meters but the grid size was doubled.  For this last case the number of Lagrangian points was also doubled.


\section{Results}\label{results}
\subsection{Comparing locomotion and flow with Reynolds number}
\begin{figure}[h!]
\begin{center}
$\begin{array}{c@{\hspace{.1in}}c@{\hspace{.1in}}c}
\includegraphics[scale=.5]{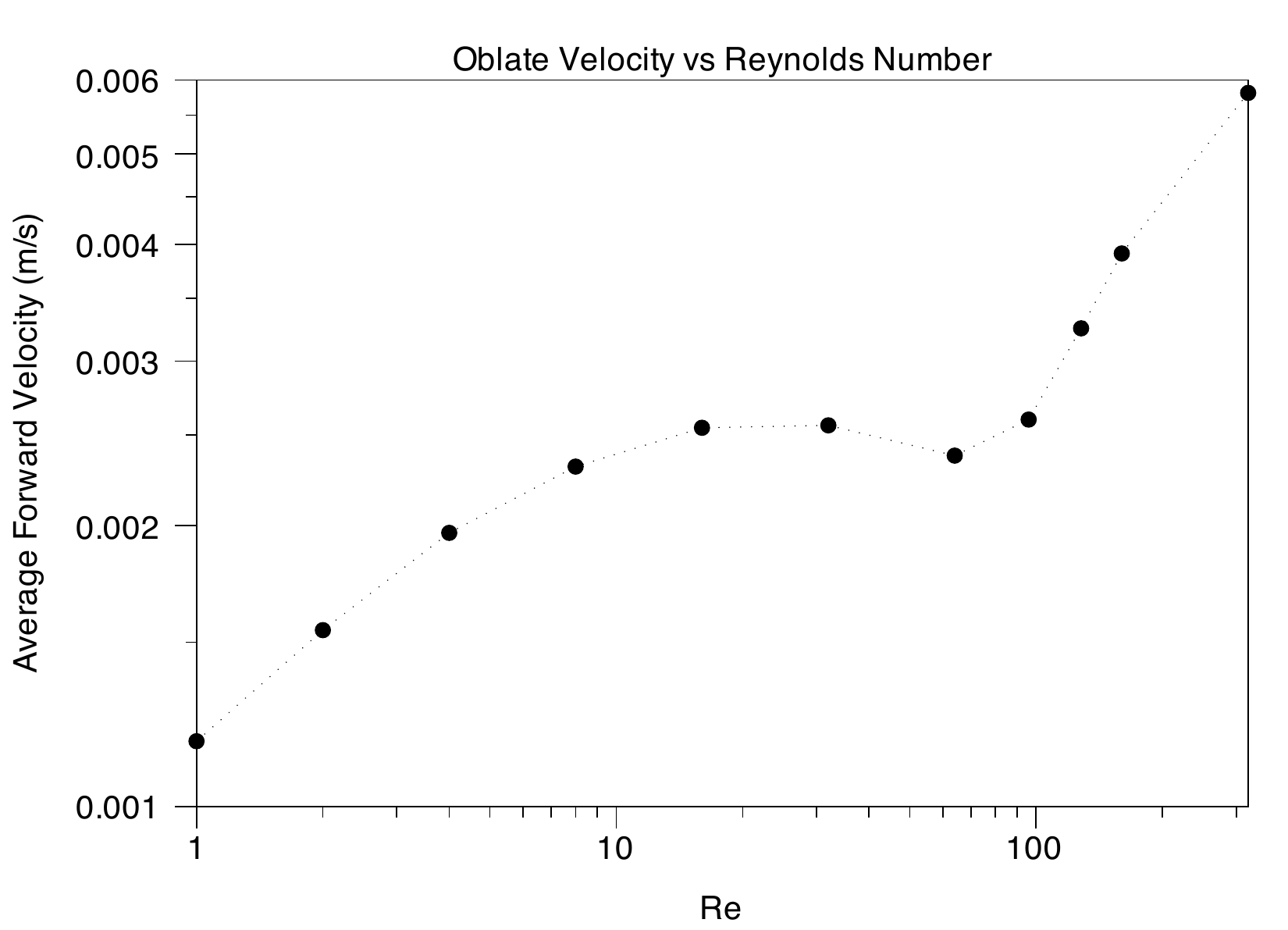} \\
\includegraphics[scale=.5]{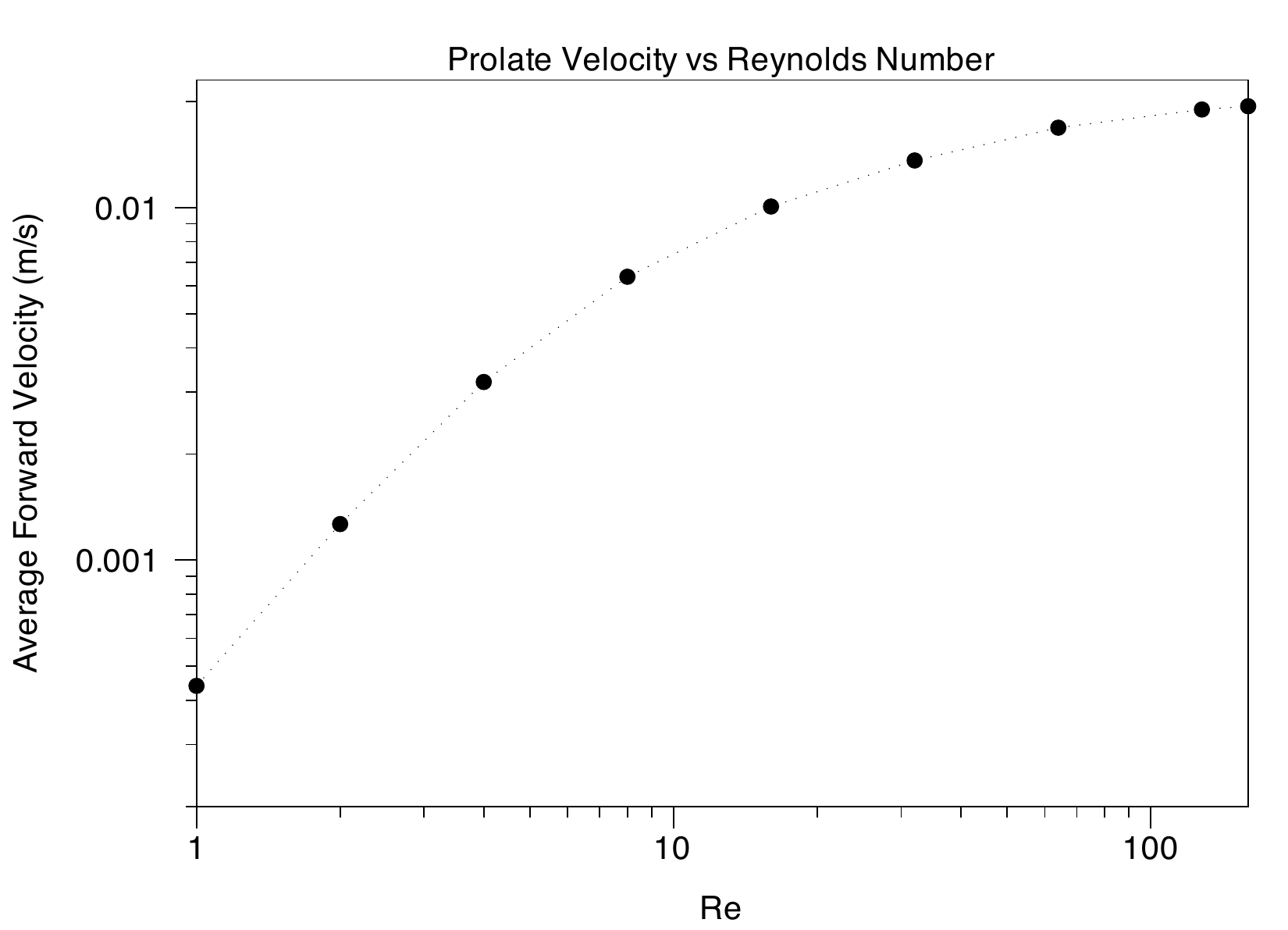} \\
\includegraphics[scale=.5]{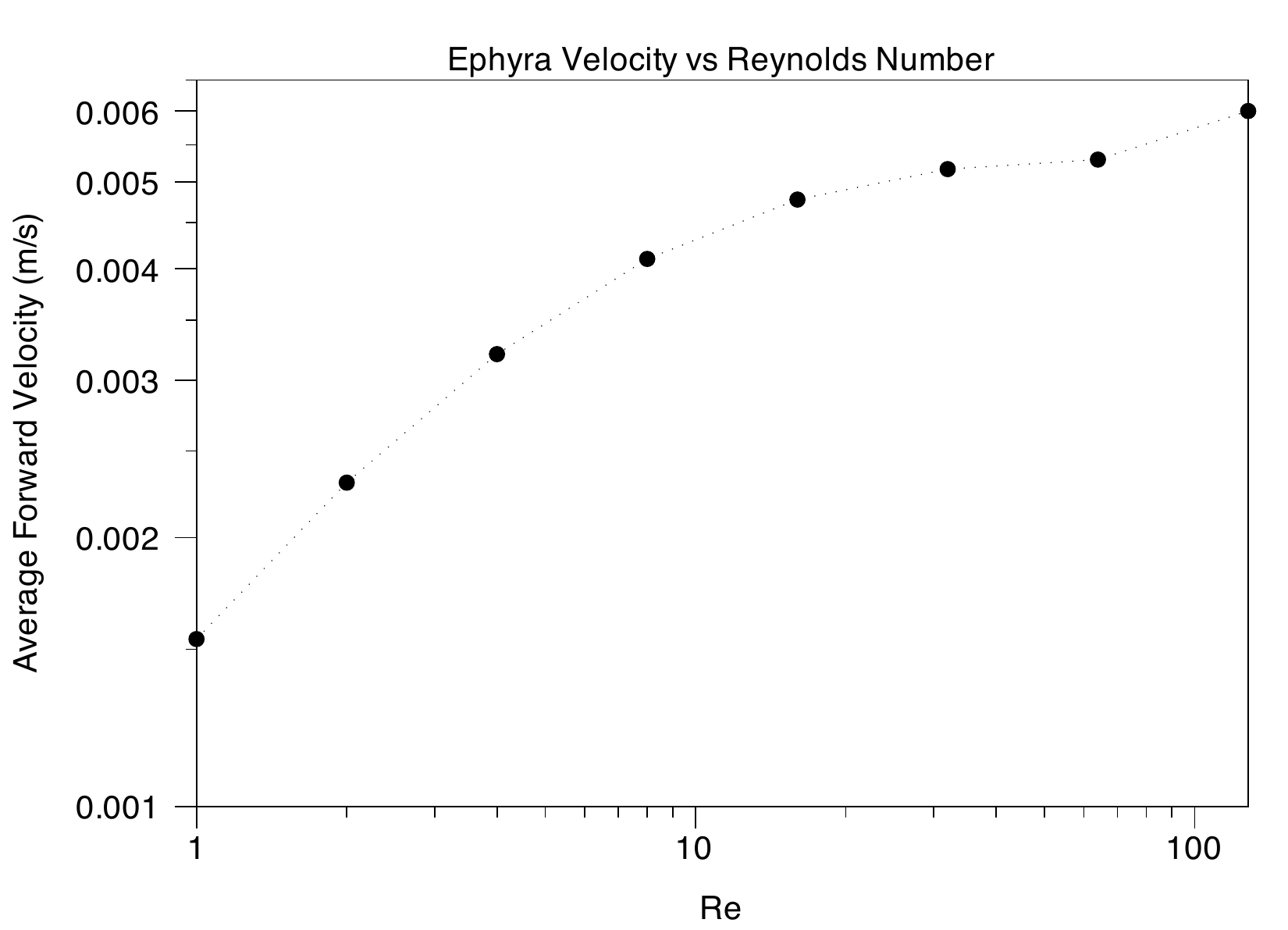} \\
\end{array}$
\end{center}
\caption{Average forward velocity as a function of $Re$ for the oblate and prolate medusae and ephyra.}
\label{fig:VvsRe}
\end{figure}

Figure~\ref{fig:VvsRe} shows the resulting average forward velocities for the oblate and prolate medusae and the ephyrae as \textit{Re} is varied. As \textit{Re} approaches zero, the average forward velocity also approaches zero. This is consistent with the predictions of the Scallop Theorem that states that reciprocal methods of locomotion will not work in Stokes flow ($Re \thickapprox 0$)~\cite{Purc:77}. In all cases, there is a significant decrease in forward velocity for $Re < 20$. For $Re>100$, the average forward velocity begins to plateau. Movies of the simulations for prolate and oblate medusae and ephyra over a range of \textit{Re} are presented in the supplemental materials. For $Re<30$, vortices are not shed from the bell margins for any of the morphologies. Vorticity quickly dissipates, especially for lower \textit{Re}. For prolate jellyfish at $Re>30$, vortices separate from the bell margin and are swept downstream, forming wakes similar to those seen in~\cite{Dab:06}. For oblate jellyfish at $Re>30$, vortices are also shed from the bell margin but circulate within and around the bell.

\begin{figure}[h!]
\begin{center}
\includegraphics[scale=.33]{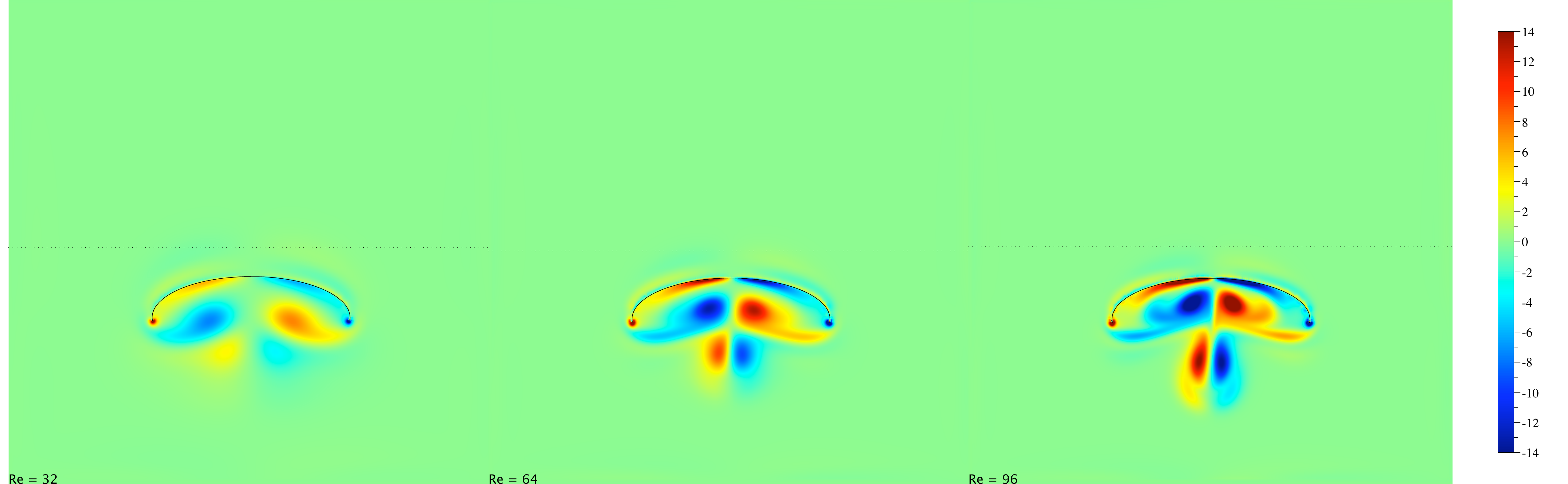}
\end{center}
\caption{The end of the second expansion is shown for the oblate jelly of \textit{Re} 32, 64, and 96.  The dotted lines show how far the medusae will travel after four full puslse cycles.}
\label{fig:oblate_Re}
\end{figure}

It is interesting to note that there is a range of \textit{Re} in which decreasing viscosity actually decreases average forward velocity for the oblate jellyfish.  This effect appears at \textit{Re} between 16 and 96.  In the the graph for the ephyra a similar effect is seen, however it is not clear whether or not the function ever truly decreases. To explain this dip, we examine the differences in flow for this range of \textit{Re}.  Figure~\ref{fig:oblate_Re} shows the vorticity patterns for \textit{Re} 32, 64, and 96.  In all cases the vortices generated by bell expansion travel back inside the bell.  This behavior becomes more exaggerated as \textit{Re} increases.  Higher $Re$ simulations show stronger and longer lasting vortices.  Much of the vorticity is pulled into the bell, and this effect grows with $Re$. Furthermore, as \textit{Re} increases the shed vortex strength that travels down stream also increases.  We propose that it is the interaction of the starting and stopping vortices (generated during contraction and expansion) that leads to this dip in forward velocity.

To evaluate the amount of work necessary to generate the contraction and expansion kinematics, the total work done, $W$, was calculated as
\begin{equation}
W = \sum\limits_{i=1}^T \sum\limits_{r=1}^N \textbf{f}_{r,i} \triangle l dx_{r,i}
\label{workin}
\end{equation}
where $\textbf{f}_{r,i}$ is the force per unit length at discrete time $i$ and Lagrangian boundary point $r$, and $dx_{r,i}$ is the distance traveled by the boundary point $r$ at time $i$.

\begin{figure}[h!]
\begin{center}
$\begin{array}{c@{\hspace{.1in}}c@{\hspace{.1in}}c}
\includegraphics[scale=.5]{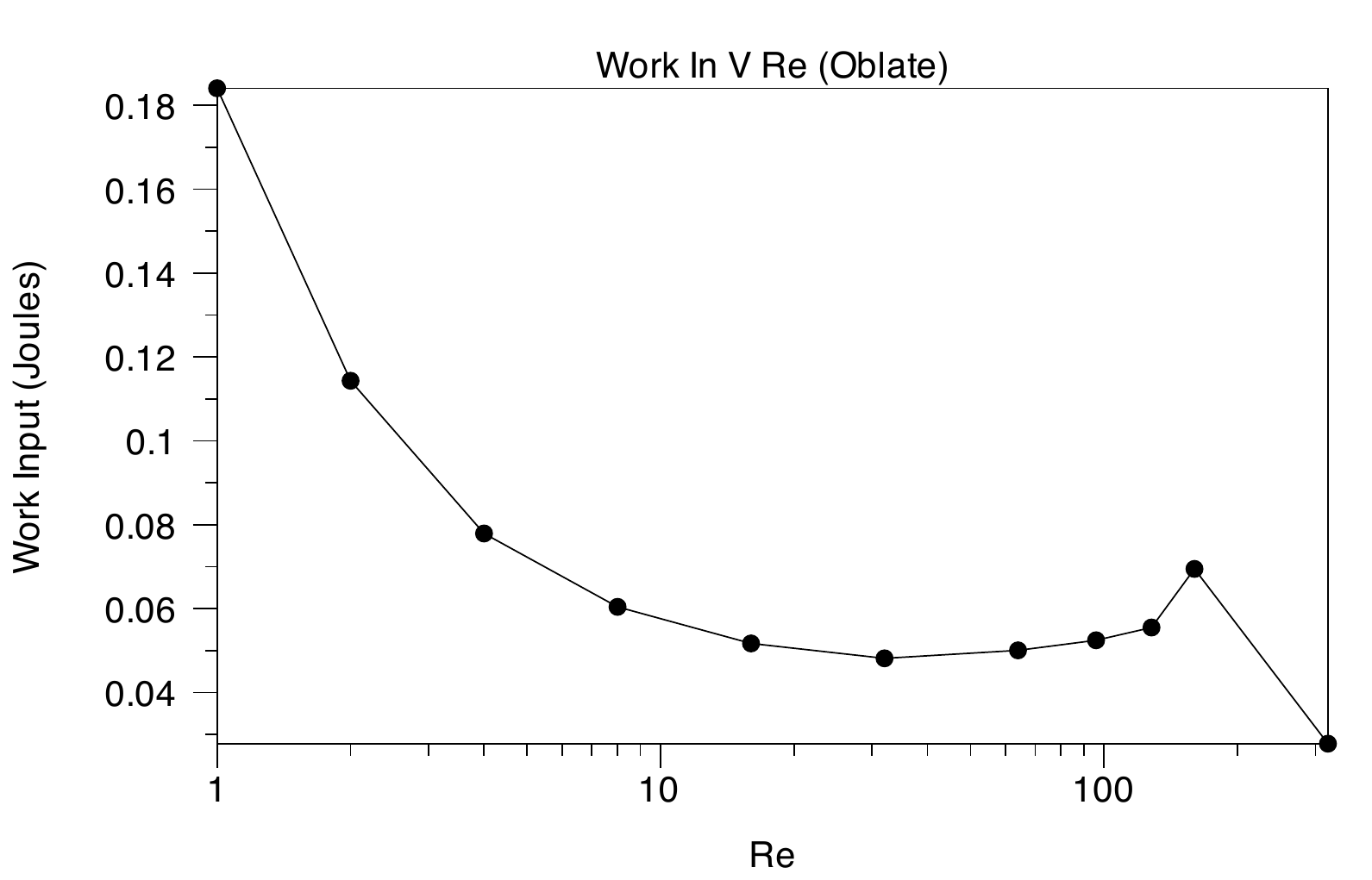} \\
\includegraphics[scale=.5]{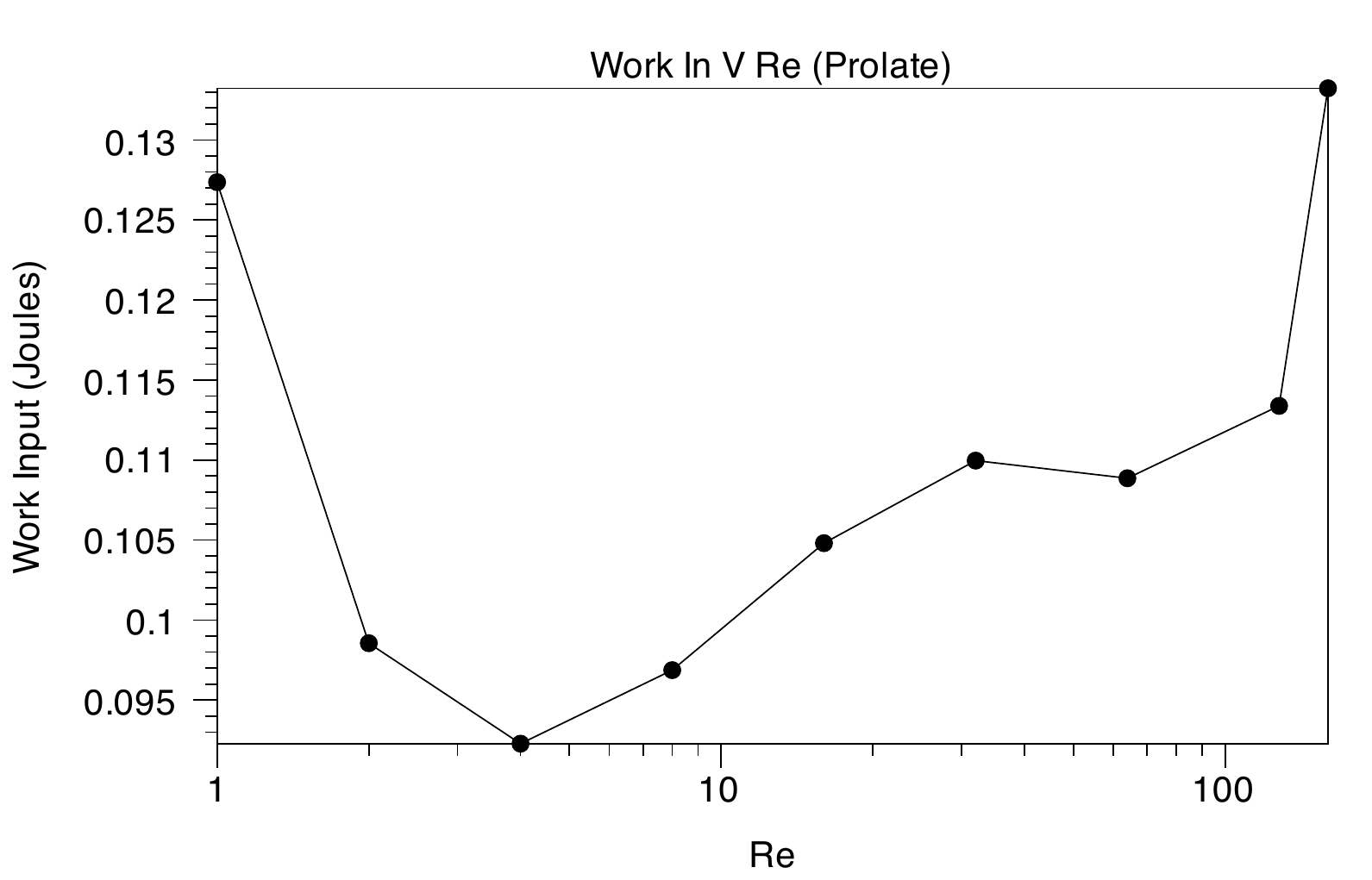} \\
\includegraphics[scale=.5]{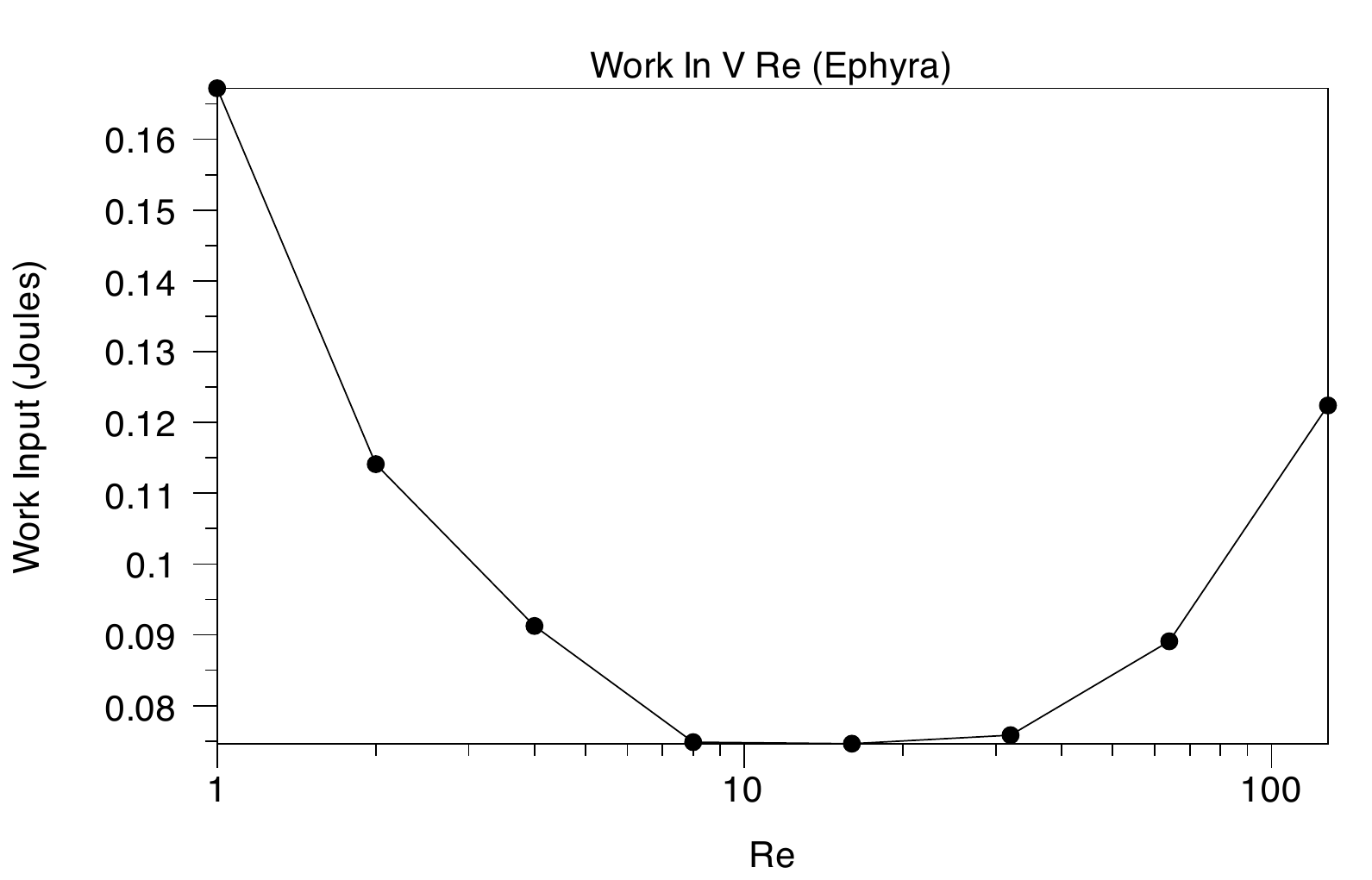} \\
\end{array}$
\end{center}
\caption{Total work done during four pulsing cycles for oblate jellyfish, ephyra, and prolate jellyfish as a function of $Re$.}
\label{fig:work}
\end{figure}

The total work done over four cycles is plotted in Figure~\ref{fig:work} as functions of \textit{Re} for the ephyra, oblate, and prolate medusae. Note that the work put into the pulsations locally peaks in all cases as the \textit{Re} approaches 1. For the oblate and ephyra cases, the work decreases as the \textit{Re} increases to 40 for the oblate jellyfish and to 10 for the ephyra. Above these values, the total work done slowly increases with \textit{Re}. For the oblate case we see a dip in work as viscosity continues to decay for $Re=320$.  For the prolate jellyfish, the total work is minimized at $Re = 4$; it then rapidly increases until $Re=32$, and finally plateaus with only small deviations until there is a large jump at $Re = 160$. Given the fact that forward velocity is also very low for $Re<10$, paddling and jet propulsion do not appear to be particularly effective mechanisms of locomotion in this range.


\begin{figure}
\begin{center}
\includegraphics[scale=.5]{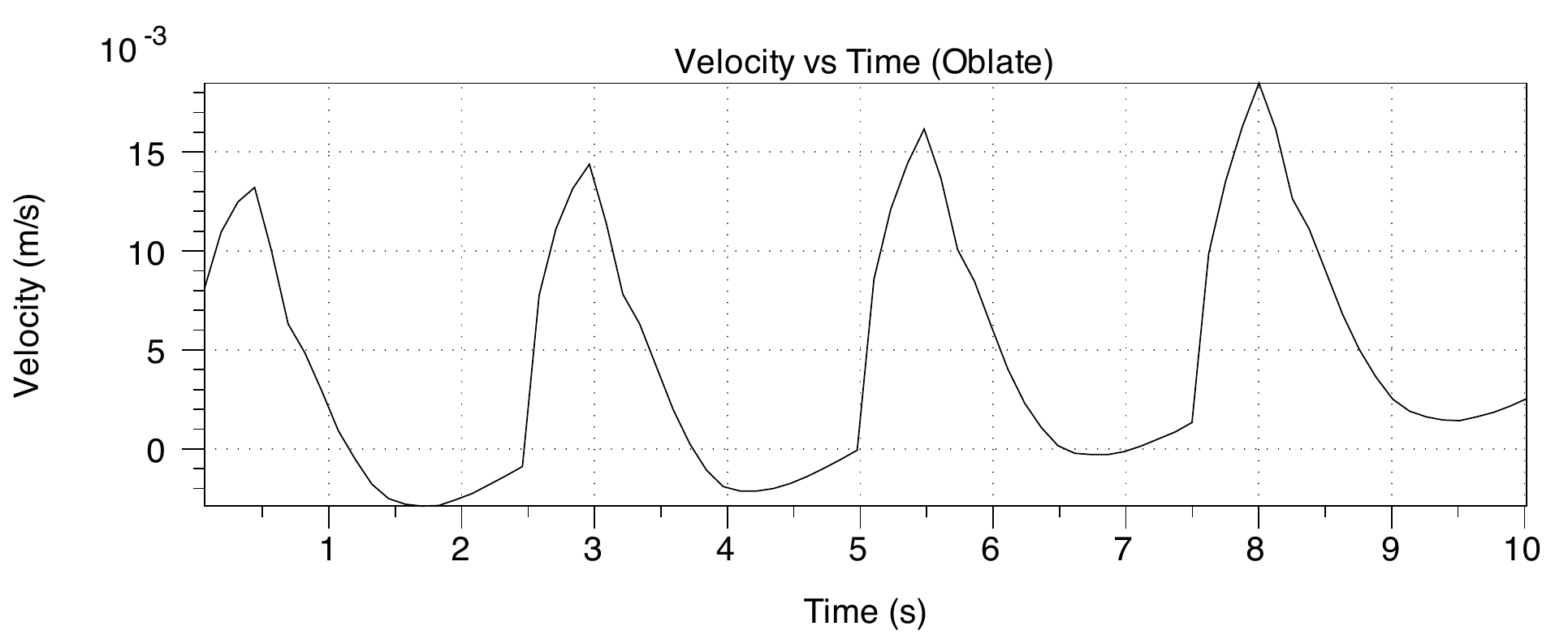}
\includegraphics[scale=.5]{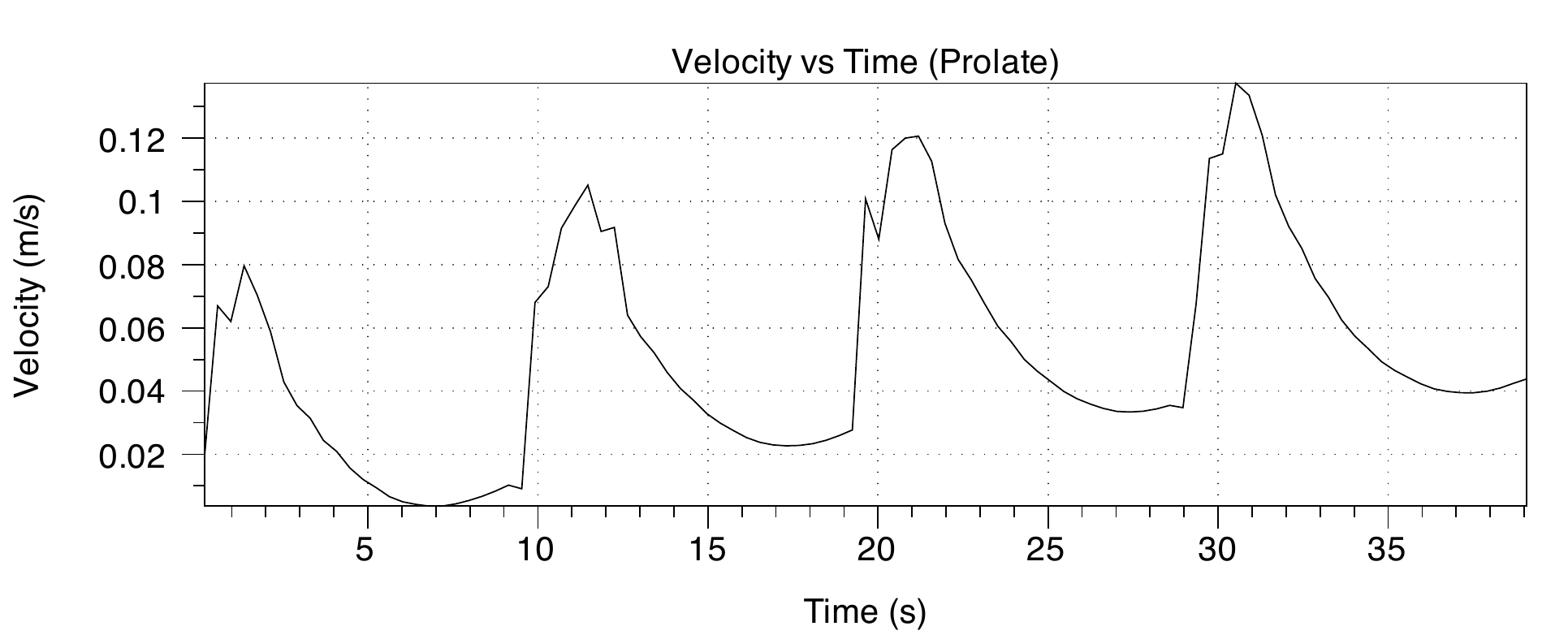}
\includegraphics[scale=.5]{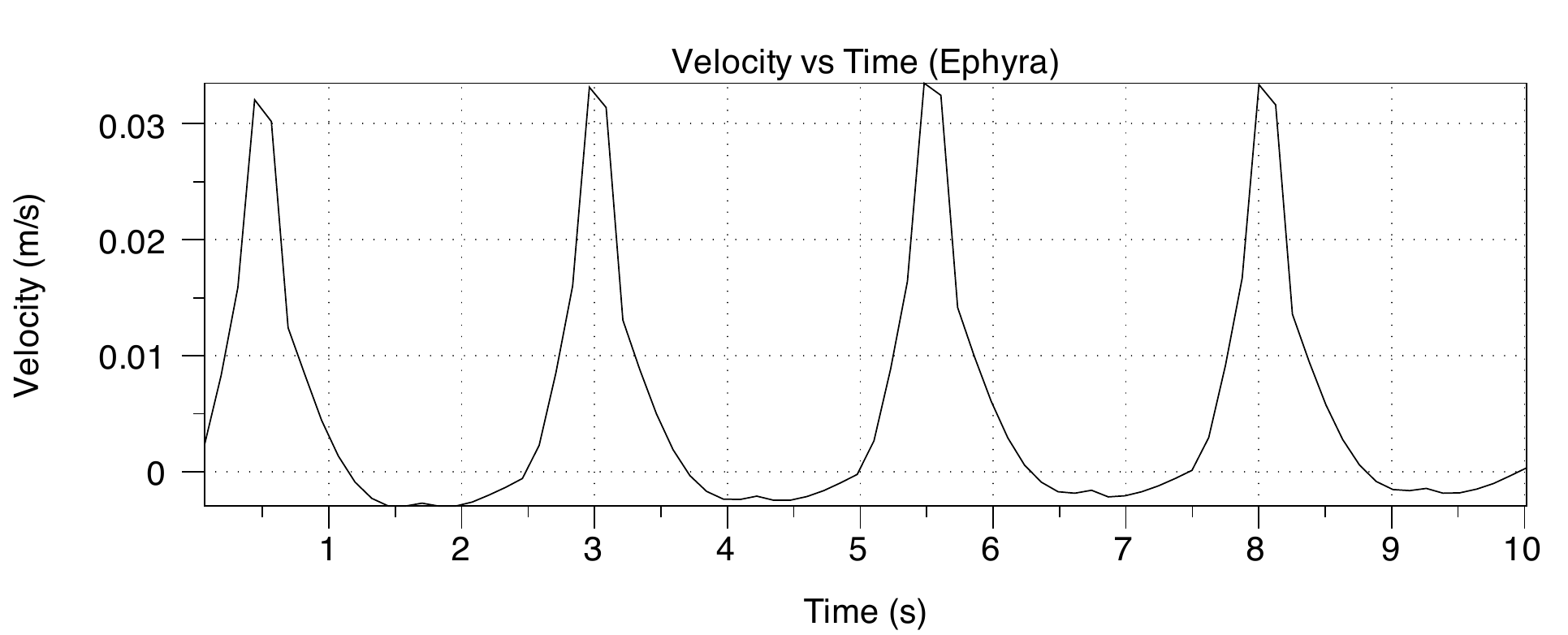}
\end{center}
\caption{Forward velocity as a function of time for the oblate, prolate and ephyra over four pulsing cycles ($Re=160, 64, 32$ respectively). Note that the velocities become negative for the cases of the oblate medusa and the ephyra. After the first cycle, prolate medusa maintain continuous forward velocities.}
\label{fig:vel_time}
\end{figure}

\subsection{Forward Velocity for Different Bell Shapes}

The  forward velocities as functions of time for the oblate and prolate medusae and ephyral cases are shown in Figure~\ref{fig:vel_time} at $Re=160, 64$ and 32 respectively. These values are chosen because they are representative of biological values (see section \ref{sec:comptojelly} below).  The jellyfish moves backward during the refilling phases in ephyral cases, and moves backwards in the oblate case during the first three refilling phases. This can be seen by the fact that the velocity becomes negative toward the end of expansion.  In all cases, the jellyfish shows significant slowing during expansion.

\begin{figure}
\includegraphics[scale=.36]{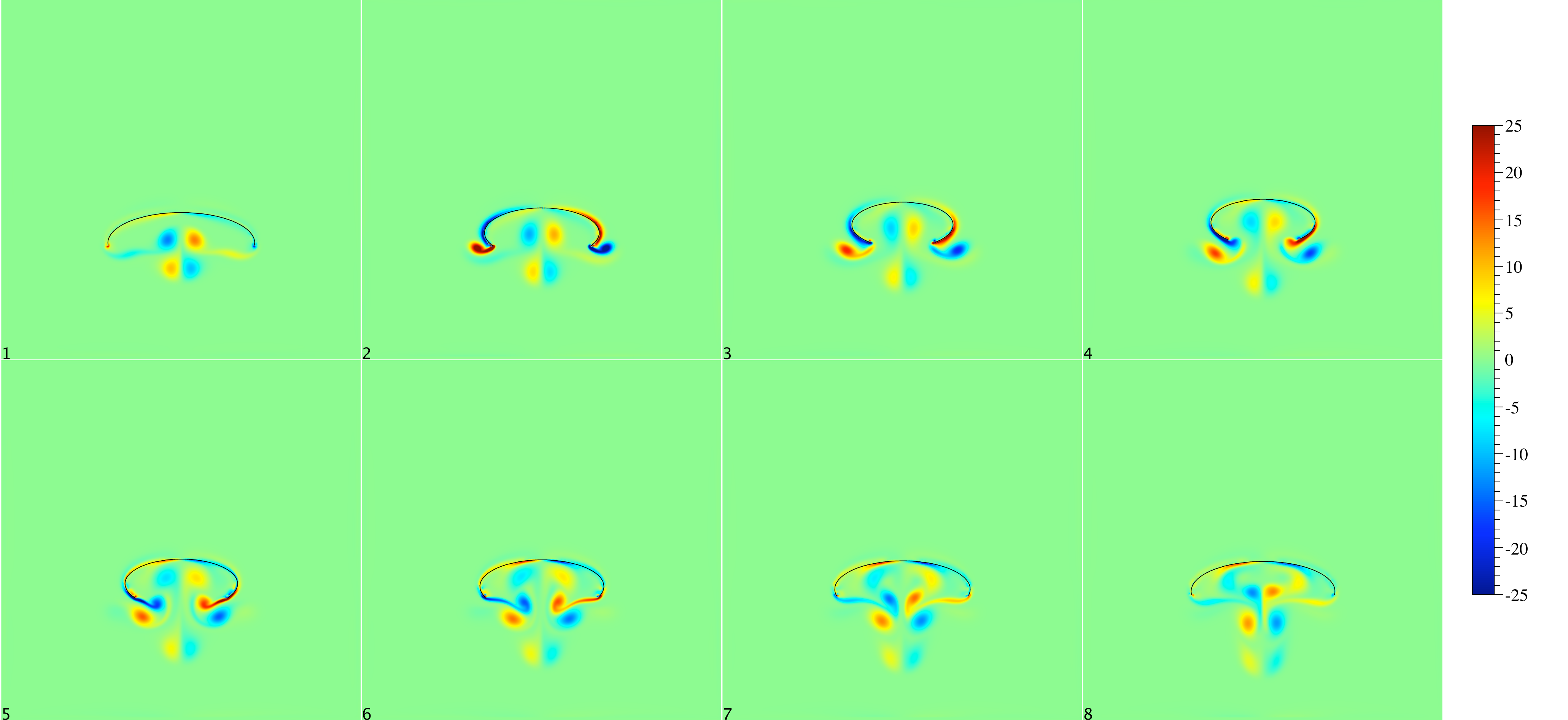}
\caption{Vorticity plots of the fluid motion around an oblate jellyfish at $Re=160$ during the second contraction. The vorticity has units of $seconds^{-1}$. Positive vorticity denotes clockwise motion while negative vorticity denotes counterclockwise motion. Frames 1-3 show the full contraction phase, and frames 4-8 show the refilling or expansion phase (at 16, 33, 50, 66, and 83 percent of the time taken to expand). Note that the bell is fully expanded in frame 1 and fully contracted in frame 3. The vorticity patterns found in frame 1 are due to the previous contraction.}
\label{fig:oblate_vort2}
\end{figure}

To gain insight into the flow structure causing the results of Figure~\ref{fig:vel_time}, time sequences of vorticity generated during contraction for the oblate medusa at $Re=160$ are presented in Figure~\ref{fig:oblate_vort2}.  Frames 4-8 show that the two shed vortices generated by contraction are only propelled gently downstream and are primarily propelled toward the center line of the body.  Frame 8 demonstrates that the generated vortices of the first contraction are pushed less than two full body lengths away from the jellyfish after nearly two full contractions. Furthermore, the two vortices formed in the first refilling phase are sucked entirely into the body, preventing another significant source of momentum from being advected downstream.  The vortices that are formed during expansion are significantly deformed by the lingering contraction vortices.  By then end of the second expansion the newly formed expansion vortices are in a similar position to those formed after the first expansion, and similarly will be largely sucked back into the body after the third contraction (see the supplementary material online).

\begin{figure}
\includegraphics[scale=.42]{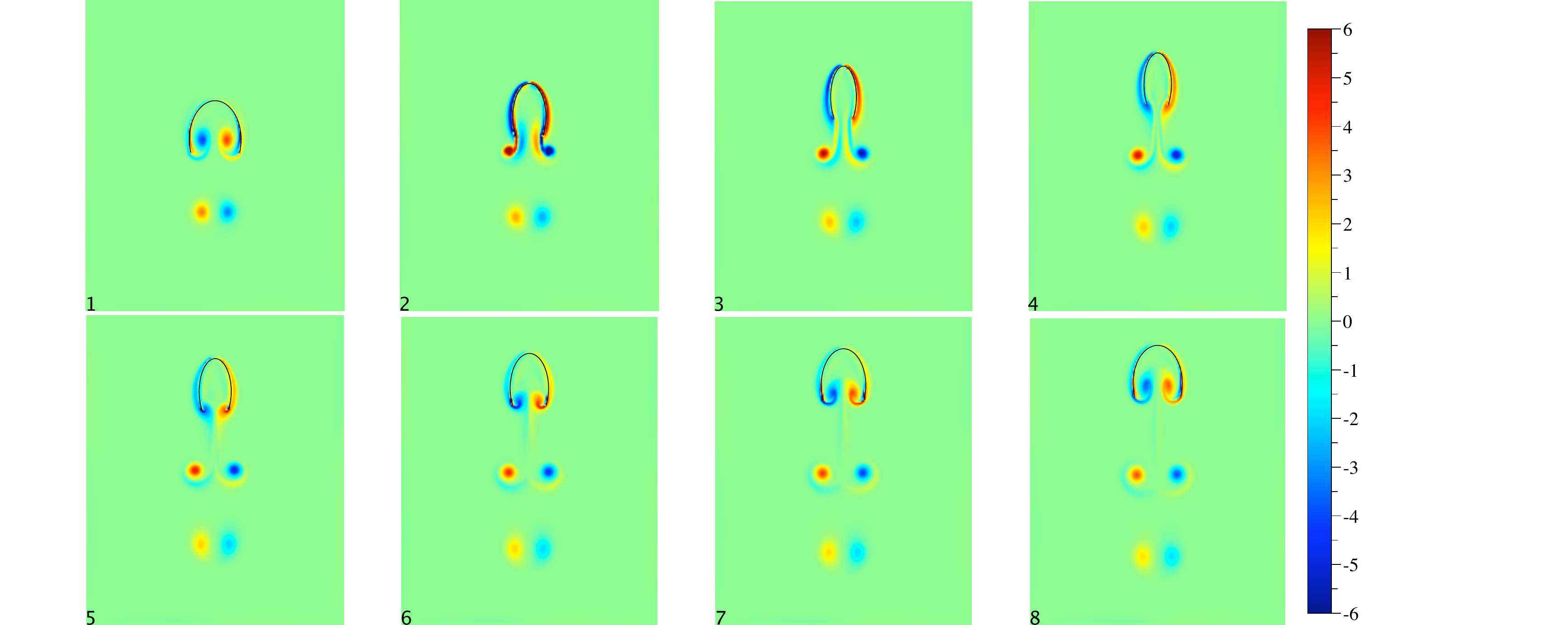}
\caption{Vorticity plots of the fluid motion around an prolate jellyfish at $Re=64$ during the second contraction. The vorticity has units of $seconds^{-1}$. Positive vorticity denotes clockwise motion while negative vorticity denotes counterclockwise motion. Frames 1-3 show the full contraction phase, and frames 4-8 show the refilling or expansion phase (at 16, 33, 50, 66, and 83 percent of the time taken to expand). Note that the bell is fully expanded in frame 1 and fully contracted in frame 3. The vorticity patterns found in frame 1 are due to the previous contraction.}
\label{fig:prolate_vort}
\end{figure}

\begin{figure}
\includegraphics[scale=.5]{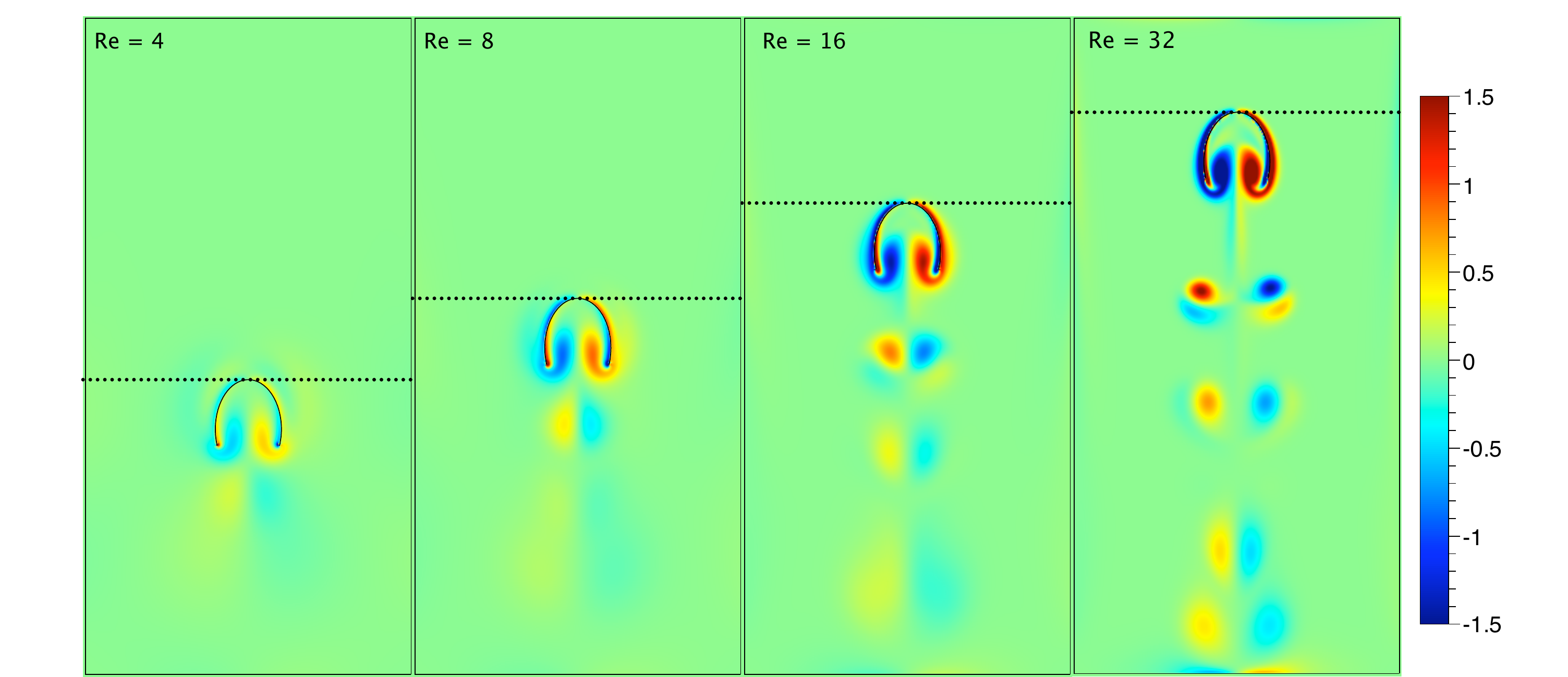}
\caption{Vorticity plots of the fluid motion around an prolate jellyfish at $Re=4,16,32,64$. The vorticity has units of $seconds^{-1}$. Positive vorticity denotes clockwise motion while negative vorticity denotes counterclockwise motion. }
\label{fig:prolate_comp}
\end{figure}

Vorticity plots for the prolate case at $Re=64$ are given in Figure~\ref{fig:prolate_vort} and show the motion over the second pulsing cycle.  The vortices that have formed after the first contraction demonstrate the strong advection of vorticity downstream.  The location of the vortices generated by expansion have been sucked back into the body, however they are largely expelled upon contraction (slides 1-3).  Furthermore, we note that the vortices generated via contraction have shed well before the end of the contraction cycle.
Note that in frames 5, 6 and 7, the closer set of shed vortices contain both the  vortices formed by contraction and those generated by expansion. Figure \ref{fig:prolate_comp} shows vorticity plots of the swimming prolate jellyfish after four contractions for $Re$ = 4, 8, 16, and 32.  Vorticity in the wake of the jellyfish quickly dissipates for lower $Re$. In addition, the separation between the body and the generated vortices decreases with $Re$.

\begin{figure}
\includegraphics[scale=.36]{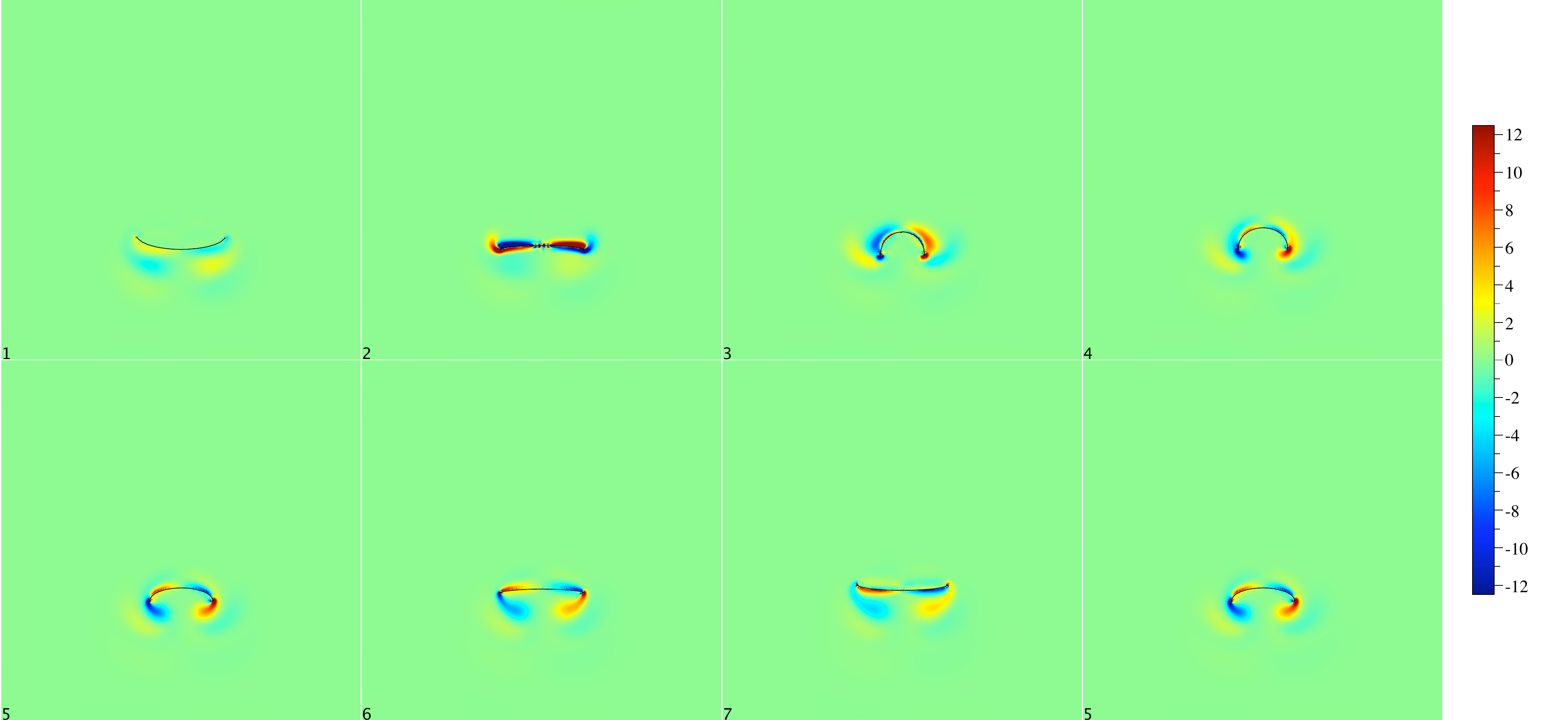}
\caption{Vorticity plots of the fluid motion around an ephyra at $Re=32$ during the second contraction. The vorticity has units of $seconds^{-1}$. Positive vorticity denotes clockwise motion while negative vorticity denotes counterclockwise motion. Frames 1-3 show the full contraction phase, and frames 4-8 show the refilling or expansion phase (at 16, 33, 50, 66, and 83 percent of the time taken to expand). Note that the bell is fully expanded in frame 1 and fully contracted in frame 3. The vorticity patterns found in frame 1 are due to the previous contraction.}
\label{fig:ephyra_vort}
\end{figure}

Figure~\ref{fig:ephyra_vort} shows vorticity plots for a time sequence of ephyral pulsing at $Re = 32$. The model ephyra displays an inability to shed vortices in a manner that effectively produces forward motion. This is easiest to see during refilling (frames 6,7,8) when the jellyfish moves backwards toward the end of expansion.  The pair of vortices that form at the bell margin during contraction do not maintain a concentrated core and are not advected downstream.  Furthermore the vortices formed in earlier contractions and expansions have deteriorated significantly.


\section{Conclusions}\label{conclusions}

\subsection{Implications for $Re$ limits of locomotion}
These 2D simplified models provide us with a first approximation of how intermediate $Re$ and geometry affect forward velocities in jet propulsion and paddling.  By solving the fluid-structure interaction problem in 2D, we are able to explore a wide parameter space. This enables us to analyze the flow patterns that form under conditions within and beyond the biologically relevant range and to explore the role of mechanical constraints on propulsive mechanisms. These simulations also allow us to develop insights on how the jellyfish might use flow patterns to enhance motion and how these patterns break down with increasing viscosity.

The numerical results in this paper show that the average forward velocities for idealized oblate and prolate medusae and ephyrae quickly approaches zero for $Re<10$. This corresponds to the lower \textit{Re} limit observed for medusae~\cite{Harg:10}, juvenile squid~\cite{Thom:01}, and octopods~\cite{Bole:01}. An examination of the vorticity plots shows that for $Re<30$, vortices do not separate from the bell margins, are not advected downstream, and as a consequence reduce the average forward velocity of the jellyfish. For $Re<4$, vorticity quickly dissipates at the end of each contraction and expansion of the bell margin.  We also find the counter intuitive result that motion can be temporarily impeded as viscosity decreases despite having over all trends of enhanced locomotion.

This work supports the idea that the behavior of intermediate $Re$ flows sets physical limits to the types of animals observed in nature. In addition to jet propulsion, Childress and Dudley~\cite{Childress:04} suggest that below a critical $Re$, flapping appendages can no longer generate enough thrust to swim or fly efficiently. Alben and Shelley~\cite{Alben:05} connect this transition to a change in the behavior of the vortex wake behind the flapping plate or appendage. The fluid field loses symmetry when vortices formed behind the plate are alternately shed from each side. These vortical structures push the body into motion. Through a similar mechanism, flapping flight does not occur for $Re<5$~\cite{Miller:04}. This is likely a result of the fact that flapping performance (defined as the ratio of lift to drag) drastically decreases for $Re<10$~\cite{Miller:09}. This drop in performance can also be connected to the behavior of vortex wake. For these low $Re$, leading and trailing edge vortices are diffuse and do not separate from the wings. Similar vortex dynamics are observed around the bell margins for the lower $Re$ cases with low swimming velocities.

Intermediate $Re$ flows also define an upper limit to ciliary and flagellar locomotion. Below some critical $Re$, animals switch from reciprocal methods of locomotion to ciliary and flagellar  locomotion.  In a very interesting case, Childress and Dudley~\cite{Childress:04} describe how the Antarctic pteropod uses ciliary locomotion at low speeds and flapping locomotion at high speeds. Boletzky~\cite{Bole:79} describes how juvenile squid use cilia to propel themselves out of the viscous egg sac and use jet propulsion once they are free in the water. Similarly, many filter feeders use jet propulsion and undulation to generate large scale feeding currents and use cilia to drive small scale flows near the filtering structures and mouths~\cite{South:55, Hadd:07}.

\begin{table}
\begin{tabular}{l*{3}{c}r}
Species             & $age$ & $diameter (m)$ & $Re_k$  \\
\hline
\textit{Aurelia aurita} & newly budded & 0.0036  & 16-45  \\
\textit{Aurelia aurita}    & smaller mature & 0.036  & 160  \\
\textit{Aurelia aurita}     & large mature & 0.102  & 1286 \\
\textit{Nemopsis bachei}     & mature & 0.007  & 62  \\
\end{tabular}
\caption{Estimates of $Re_k$ using the bell diameter and average bell tip velocity during contraction. }
\label{jelly_re}
\end{table}


\subsection{Comparisons to jellyfish}
\label{sec:comptojelly}
To compare the presented model with the jellyfish found in nature, we estimate $Re_k$ values of real jellyfish found in nature and the results are summarized in table \ref{jelly_re}.  The presented values for the oblate case representing a large \textit{Aurelia}  yield $Re = 1286$ (taking the kinematic viscosity of sea water to be $\nu = 1.05\times10^{-6}$m$^2$s$^{-1}$ ~\cite{bullard:10}).  The smaller mature oblate case in Dabiri \textit{et al.}~\cite{Dabi:05} is cited as having diameter of 3.6 cm with similar kinematics.  Scaling the lengths appropriately and keeping the same contraction time and precent changes yields a Reynolds number of 160.   For the prolate case we estimate $Re$ by  taking $a_i = .0035$ m, $b_i = .004$ m, $p_a = .75$, $p_b=.1$, and $t_c=.15$ s which gives $Re = 59$~\cite{Dab:06}.  Finally for the ephyral case we acknowledge that the role of $b_i$ and $p_b$ play a dominant role in the $Re$ calculation, but careful measurements of ephyral bell kinematics have not been reported in the literature.  To determine the range of $Re$ for a range of ephyrae, we estimate the upper bound of $Re$ by replacing $U_{body}$ with the maximum tip velocity found in Feitl ~\cite{Feitl:09}, which is 25mm s$^{-1}$, and take a lower bound by setting $p_b=0$.  Estimating $a_i=.18$ cm, $p_a=.5$, and $t_c=.1$ s ~\cite{Feitl:09}, we find that a reasonable range of $Re$ for the ephyrae is between 16 and 45.  With these estimates we choose the corresponding closest $Re$ from our simulations and compare it to the swimming dynamics of ephyrae.

\begin{figure}
\begin{center}
\includegraphics[scale=.7]{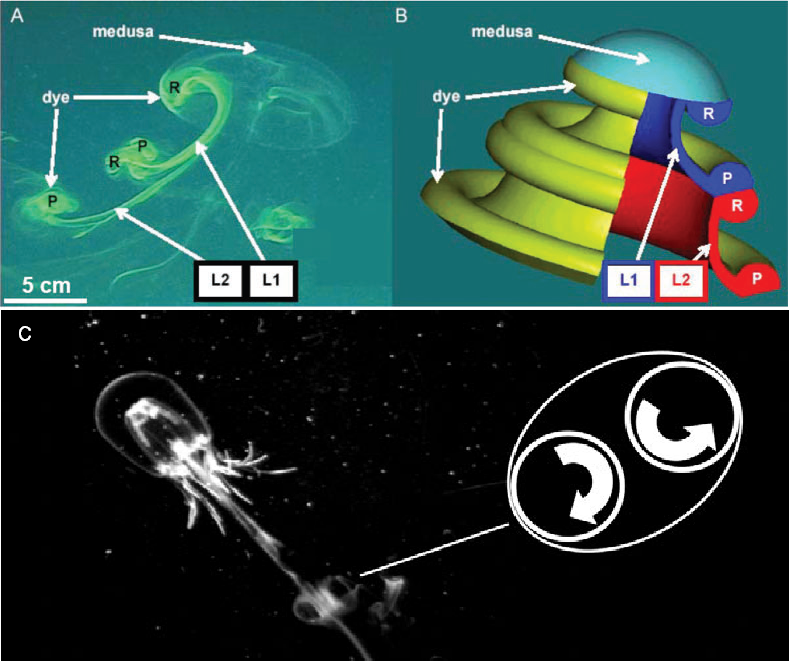}
\end{center}
\caption{(A) Flow visualization of the wake of the oblate moon jellyfish \textit{Aurelia aurita} from Dabiri \textit{et al}~\cite{Dabi:05}. (B) Corresponding schematic diagram of the vortex wake. P shows the vortex ring formed during contraction, R shows the vortex ring generated during bell expansion, and L1/L2 are label for the adjacent lateral vortex superstructures. (C) Flow visualization of vortex formation in the prolate jellyfish \textit{N. bachei} from Dabiri \textit{et al.}~\cite{Dab:06}. The starting vortex generated during contraction is rapidly swept downstream of the jellyfish.}
\label{fig:real_jellies}
\end{figure}

After two contractions, the oblate jellyfish at $Re=160$ moves 80 percent of its body length per contraction. This is comparable to steady state swimming velocities measured for \textit{Aurelia} that move a little over a body length per contraction~\cite{Costello:94, dabirivid:09}. Note, however, that the oblate model does not move as far as smaller mature \textit{Aurelia aurita} from rest; the model moves only two body lengths over four contractions.  To understand the differences in performance between the real and model oblate jellyfish, we examine the vortex dynamics.   Dabiri \textit{et al.}~\cite{Dabi:05} describe how the starting vortex ring generated during contraction and the stopping vortex generated during expansion separate from the bell margin and travel away from the bell together. This pair of starting and stopping vortex rings form a lateral vortex superstructure (see Figure~\ref{fig:real_jellies}A,B). Adjacent lateral vortex superstructures then pull fluid into the wake and downstream of the jellyfish, enhancing forward propulsion.  In contrast to this structure, the starting and stopping vortices generated by the model jellyfish are pulled back into the body, even during the final contraction when the jellyfish is moving the fastest (see the supplemental material).  Furthermore, the vortices generated by the contraction of the model oblate jellyfish collapse toward the centerline, unlike the vortex dynamics observed in actual \textit{Aurelia}~\cite{dabirivid:09}. For certain parameter values, the model oblate jellyfish moves with negative velocity during the expansion phase. This phenomenon has been experimentally verified in many oblate species such as \emph{Mitrocoma cellularia}~\cite{Coli:02}, \emph{Cyanea capillata}~\cite{Colin:95}, and \emph{Aurelia aurita}~\cite{Costello:94}.

The differences between the vortex dynamics of the oblate model and the real jellyfish may originate from a combination of 2D effects and from the kinematic stiffness that is imposed.
The 2D simulations imply that the generated vortices act as rigid cylinders whose cross sections translate within the 2D plane.  Once shed during the contraction phase, it is expected that the vortices will move at the local fluid velocity in the 2D plane with no vortex stretching~\cite{Ache}.  In the 3D case however, one vortex ring is generated during each contraction and each expansion.  In order for the vorticity to move toward the centerline, the single ring would have to shrink with vorticity concentrated in a small volume. In reality, the vortex rings generated by oblate jellyfish stretch rather than shrink.  The ability of the vortices to remain a short distance from the bell margin may play a crucial role in the complex starting and stopping vortex interactions found in ~\cite{Dabi:05}.  It is reasonable to suppose that if the vortices where not pulled into the centerline then the generated stopping (expansion) vortex would help push both rings downstream.  In the 2D case, the momentum of the starting (contraction) vortices pushes the stopping vortices back into the bell.  The model also employs a stiff paddling motion of the lower bell which appears to sweep the vortex structures generated during expansion inside the bell.  In contrast, the actual jellyfish has flexible tissue called the vellum which may allow for the vortex structures to advect downstream rather than becoming trapped inside the bell.




In the case of the prolate medusae, such as \textit{Nemopsis bachei}, a single vortex ring is generated during the contraction that is quickly swept downstream~\cite{Dab:06} (see Figure~\ref{fig:real_jellies}C). The complicated vortex superstructures consisting of oppositely spinning starting and stopping vortices observed in oblate medusae are not present for the prolate species. We find the same pattern of vortex formation and separation for the model prolate jellyfish and, in both cases, vortices generated via contraction have shed before the end of the contraction cycle. Due to the conservation of total vorticity~\cite{Wu:81}, oppositely signed vorticity is generated during bell expansion, but coherent stopping vortex structures do not form. It appears that the 2D approximation does a reasonable job of capturing the vortex dynamics for the prolate jellyfish. This may be aided by the fact that complicated 3D interactions between starting and stopping vortex rings do not need to be resolved to obtain reasonable dynamics. In the simulations, the prolate at $Re$ of 32 and higher move between five and eight body lengths in four contractions after starting from rest, which is similar to what is observed by Dabiri ~\cite{Dab:06}. For $Re=64$, the model prolate moves roughly six body lengths over four contractions which is slightly larger than what is observed in \textit{N. bachei}.  We note however that the model prolate is a slender body without the appendages and body girth found in nature that would introduce added drag.

As for the ephyral case, we note that to our knowledge particle image velocimetry (PIV) studies have not yet been carried out on the living organism for quantitative comparison. Feitl \textit{et al.}~\cite{Feitl:09} visualized the flow around free swimming ephyrae. They note that predominance of viscous forces and describe how circulating vortices are not created in the wake. This is similar to the results found in this study for lower $Re$.  Similarly, there have not been rigorous visualizations of vortex formation and shedding for ephyra. Nawroth \textit{et al.}~\cite{Nawroth:10} found that the ephyrae of \textit{Aurelia aurita} travel 0.5 - 1 body length per pulsation cycle, which is higher than the swimming speeds measured for the model ephyrae. It seems likely that 3D effects as well as the complex structure of the bell play a role in swimming performance for the juvenile jellyfish.

\subsection{Limitations and future work}
The goal of this study was to explore the performance of simplified organisms using jet propulsion over a range of intermediate $Re$. Two-dimensional simulations of jetting modeled by time varying deformations of a hemiellipsoid allow a large parameter space of shapes, kinematics, and $Re$ to be explored. These simulations clearly show a sharp drop off in swimming performance as $Re$ decrease below about 10, but note that the work here is not a substitute for the careful modeling of specific animals such as the work of~\cite{Mohs:09,Sahi:09,Lipi:09}. The simplified model matches well with body lengths traveled per contraction and qualitatively with the prolate case but does not capture the complex vortex dynamics observed in oblate jellyfish. Future work to improve the model for a targeted study could be made by adding flexibility to the jellyfish, incorporating three-dimensional effects with an axisymmetric solver, carefully modeling the morphology and contraction kinematics for a specific species of jellyfish, etc.

It is also worthwhile to note that simulations using Peskin's standard immersed boundary method~\cite{Peskin02} to resolve the shedding of vortex structures off these sharp boundaries for $Re>1000$ requires spatial grid sizes that are prohibitively small. As such the authors were not able to explore the full range of $Re$ for which organisms that use jet propulsion live.  Alternative methods that better handle sharp boundaries at higher $Re$ would be valuable for exploring the dynamics of jet propulsion at these larger scales.


\section{Acknowledgements}
We would like to thank Christina Hamlet, Terry Campbell, William Kier, Ty Hedrick, and Arvind Santhanakrishnan for their advice and insights throughout this research project. This work was partially funded by the Burroughs Wellcome Fund (BWF CASI ID\# 1005782.01) and by the National Science Foundation (NSF FRG \#0854961).

\bibliographystyle{plain}

\bibliography{jelly_refs}

\end{document}